\documentclass[12pt]{article}

\usepackage[a4paper,text={16.8cm,22.4cm}]{geometry}
\PassOptionsToPackage{dvipsnames}{xcolor}
\usepackage{hf-tikz,amsmath,amsfonts,braket,slashed,amssymb,bm,psfrag,graphicx,color,colortbl,dsfont,euscript,ctable,bbm,hyperref} 
\usepackage[small,labelfont=bf]{caption}
\usetikzlibrary{arrows.meta}
\RequirePackage[sort&compress,square,comma,numbers]{natbib}
\usepackage{stmaryrd} 
\usepackage{comment}
\usepackage{soul}
\allowdisplaybreaks
\renewcommand{\arraystretch}{1.3}

\numberwithin{equation}{section}

\newcommand{\vsl}{\rlap{\hspace{0.25mm}/}{v}}
\newcommand{\nsl}{\rlap{\hspace{0.25mm}/}{n}}

\newcommand{\nbsl}{\rlap{\hspace{0.25mm}/}{\nb}}

\newcommand{\dilog}[1]{\operatorname{Li}_2\left( #1 \right)}

\newcommand{\nb}{{\bar n}}

\newcommand{\abs}[1]{\left| #1 \right|}

\newcommand{\gdirac}[2]{#1 \otimes #2 }
\newcommand{\barn}{{\bar n}}

\newcommand{\vl}{{v_\tau}}

\def\A{{\EuScript A}}
\def\F{{\EuScript F}}

\newcommand{\spac}{{\hspace{0.3mm}}}

\definecolor{LightGray}{rgb}{0.85,0.85,0.85}

\definecolor{AccentColor}{rgb}{0.0,0.47,.90}
\newcommand{\shadecolor}[1]{AccentColor!#1!white}

\begin{document}

\begin{titlepage}

\begin{flushright}
\normalsize
MITP-26-027\\
July 14, 2026
\end{flushright}

\vspace{5mm}
\begin{center}
\LARGE\bf\boldmath
QED Corrections to $B^-\to\tau^-\spac\bar\nu_\tau$
\end{center}

\vspace{0.4cm}
\begin{center}
\textsc{Claudia Cornella$^a$, Max Ferr\'e\spac$^b$, Matthias K\"onig\spac$^b$ and Matthias Neubert\spac$^{b,c}$}

\vspace{6mm}
\textsl{${}^a
$ Istituto Nazionale di Fisica Nucleare, Sezione di Padova\\
Via F. Marzolo 8, 35131 Padova (PD), Italy
\\[0.3cm]
${}^b$PRISMA$^{++}$ Cluster of Excellence \& Mainz Institute for Theoretical Physics\\
Johannes Gutenberg University, Staudingerweg 9, D-55128 Mainz, Germany\\[0.3cm]
${}^c$Department of Physics \& LEPP, Cornell University, Ithaca, NY 14853, U.S.A.}
\end{center}

\vspace{0.8cm}
\begin{abstract}
Using a sequence of effective field theories (EFTs), we calculate the rate for the leptonic decay $B^-\to\tau^-\spac\bar\nu_\tau(\gamma)$ including real and virtual QED corrections, with a cut $E_{\rm cut}\ll\Lambda_{\rm QCD}$ imposed on electromagnetic radiation in the $B$-meson rest frame. We establish a factorization theorem for the rate and evaluate it at $\mathcal{O}(\alpha)$, resumming the leading logarithmic corrections to all orders in perturbation theory. The large mass $m_\tau$ allows us to treat the tau lepton as a heavy fermion below $\mu\sim m_B\sim m_\tau$, leading to an EFT construction that is structurally different and noticeably simpler than that for the muon case \cite{Cornella:2026lkp}. In particular, hadron-structure dependent QED corrections can be described in terms of QED-induced $B^-\to\tau^-$ form factors, which should be calculable on the lattice. Our analysis includes the leading $\Lambda_{\rm QCD}/m_\tau$ corrections as well as the leading corrections in $E_{\rm cut}/\Lambda_{\rm QCD}$. We show that contributions of the form $\Lambda_{\rm QCD}\spac m_B/m_\tau^2$, which are naively sizable, cancel among each other. Contrary to the muon channel, structure-dependent corrections involving $BB^\ast\gamma$ transitions are numerically negligible for the tau case. The remaining logarithmic dependence on $E_\mathrm{cut}$ is mild. We estimate that the present uncertainty in the calculation of the structure-dependent QED corrections is about 0.5\% of the rate. As a byproduct, we present a state-of-the-art prediction for the lepton flavor universality ratio of the tau and muon channels.
\end{abstract}

\end{titlepage}
\spac
\tableofcontents
\hspace{1cm}

\section{Introduction}

The leptonic decay $B^-\to\ell^-\bar\nu_\ell$ is one of the theoretically cleanest processes in flavor physics. When QED corrections are neglected, its decay rate is proportional to $|V_{ub}|^2 f_B^2$. This makes it a particularly clean channel for the extraction of the CKM matrix element $|V_{ub}|$, given that the $B$-meson decay constant $f_B$ is known with sub-percent precision from lattice QCD~\cite{FlavourLatticeAveragingGroupFLAG:2024oxs}. Beyond the Standard Model, the decay is sensitive to new charged-current interactions, so precise measurements also provide a valuable probe of new physics. The branching fraction $\text{Br}(B^-\to\tau^-\spac\bar\nu_\tau)$ has been measured by BaBar, Belle and Belle~II \cite{BaBar:2009wmt,BaBar:2012nus,Belle:2012egh,Belle:2015odw,Belle-II:2025ruy}, but still with large uncertainties of $\mathcal{O}(20\%)$. The ongoing Belle~II program is expected to reduce this to 5--6\%~\cite{Belle-II:2018jsg,Belle-II:2022cgf}. On a longer timescale, a future Tera-$Z$ machine such as FCC-ee could push the precision down to the percent level \cite{Zuo:2023dzn}. 

Achieving such an experimental accuracy requires a comparably precise theoretical prediction. In particular, electromagnetic corrections must be treated systematically, including the soft-photon radiation that inevitably accompanies any process involving charged particles. In a recent paper, we derived the state-of-the-art prediction for $B^-\to \ell^-\spac\bar\nu_\ell(\gamma)$ focusing on the case $\ell=\mu$ \cite{Cornella:2026lkp,Cornella:2022ubo}.  It is then natural to ask how the other lepton channels compare with the muonic one. In the electron case, the strong chiral suppression $m_e/m_B \ll 1$ renders the decay essentially unobservable. The situation is very different for the tau channel. Since the tau mass is comparatively large, the helicity suppression is much milder than in the electron or muon modes. Because of the modest mass hierarchy ($m_\tau/m_B\approx 1/3$), we argue that treating the tau lepton in direct analogy with the muon is not well justified, and that it should instead be considered as a second heavy particle. The appropriate effective description for scales between $m_B$ and the hadronic scale $\Lambda_{\rm QCD}$ is then an HQET-like theory with two heavy fermions, which we denote by HFET (heavy-fermion effective theory). This setup is reminiscent of $B\to D$ transitions, for which the charm quark is usually treated as a heavy quark. 

Our study will be performed under assumptions similar to those adopted for the muon channel. In particular, we consider a setup in which the experiment only allows for additional final-state radiation with total energy below a cut $E_\mathrm{cut}\ll\Lambda_{\rm QCD}$. For such small values of the veto energy, the allowed real radiation is soft enough to see the meson only as a point-like particle. For a radiation veto of similar or larger magnitude than $\Lambda_\mathrm{QCD}$, real radiation would probe the inner structure of the $B$ meson and could excite it into higher resonance states, in which case a theoretical description would involve unknown hadronic form factors (see e.g.\ \cite{Kurten:2022zuy}). Therefore, the upper limit on the veto chosen here is crucial for a precise treatment of the process. Virtual corrections, instead, are insensitive to this cut and can involve radiation of sufficiently short wavelength to probe the inner structure of the decaying $B$ meson. In contrast to the muon channel (and in principle also the electron channel), these corrections generate only local hadronic currents, whose matrix elements do not involve light-cone distribution amplitudes. Nevertheless, the presence of the charged final-state lepton makes these matrix elements non-universal, so that the QCD decay constant $f_B$ must be generalized to QED-induced $B^-\to\tau^-$ form factors, which can in principle be studied using lattice field theory.

At scales $\mu\sim\Lambda_{\rm QCD}$ we match onto an EFT in which the mesons are treated as point-like particles, while the leptonic degrees of freedom remain unchanged. In \cite{Cornella:2026lkp}, the appropriate theory was constructed with the inclusion of light pseudoscalar mesons, resulting in what is commonly called heavy-hadron chiral perturbation theory. As we have shown, for the analysis of leptonic $B$ decays all effects involving pions can be absorbed into the renormalization of the couplings of the heavy mesons, such as $g_{BB^\ast\gamma}$.\footnote{The only exception is the contribution $B^-\to B^{*-}\spac\pi^0\to\tau^-\spac\bar\nu_\tau\spac\gamma\gamma$, in which the on-shell soft pion decays with unit probability into two photons. To remove this background, one should either require that $E_{\rm cut}<m_{\pi^0}$ or reconstruct the invariant mass of the two photons.} 
Here we therefore omit the introduction of light meson fields and treat the low-energy dynamics in a simplified heavy-meson effective theory (HMET), which is the same as in the muon case. It treats the $B$ meson and its lowest-lying excited state $B^\ast$ as an almost mass-degenerate doublet under the heavy-quark spin symmetry. In the muon case, a numerically relevant contribution arises from the radiative transition of the decaying $B$ into a virtual $B^\ast$, which subsequently decays leptonically: $B^-\to B^{\ast-}\gamma\to\mu^-\spac\bar\nu_\mu\spac\gamma$. The power suppression of the $B$--$B^\ast$ interaction is offset by the lack of a chiral suppression $\sim m_\mu/m_B$ in the leptonic decay of the $B^\ast$ meson. This compensation mechanism is absent for the tau channel, because the mass ratio $m_\tau/m_B\approx 1/3$ does not provide a significant suppression. As a result, we find that intermediate $B^\ast$ contributions can be safely neglected.

The theoretical description of the $B^-\to\tau^-\spac\bar\nu_\tau(\gamma)$ decay rate involves a sequence of EFTs. Starting from the Low-Energy Effective Theory (LEFT) below the electroweak scale, we match at $\mu_h \sim m_B\sim m_{\tau}$ onto HFET. Subsequently, at a hadronic scale $\mu_0=1.5$\,GeV that is still in the perturbative regime, we match onto HMET. The resulting construction can be illustrated below:\\[2ex]

\begin{minipage}{\textwidth}
\begin{center}
\centering
 \def\y0{12}        
 \def\ya{10}        
 \def\yb{7}         
 \def\yt{6.5}       
 \def\yh{3.2}       
 \def\yr{4.3}       
 \def\yd{1.3}       
 \def\ye{0}         
 \def\yl{0.2}       
 \def\ylo{-1.2}     
 \def\ylabel{-1.6}  
 \def\vs{2}         
\begin{tikzpicture}
 \fill[\shadecolor{10}](\ya,0 - \vs)rectangle(\y0,1 - \vs);  
 \fill[\shadecolor{20}](\yh,0 - \vs)rectangle(\ya,1 - \vs);  
 \fill[\shadecolor{25}](\yh,0 - \vs)rectangle(\ylo,1 - \vs); 
 \draw[thick,-Latex]  (\ylo,0 - \vs) -- (\y0,0 - \vs);
 \node at (\y0,-.4 - \vs){$\mu$};

 \draw[thick] (\ya,-.1 - \vs)--(\ya,.1 - \vs); 
 \draw[thick, dotted] (\ya, 0 - \vs)--(\ya,1 - \vs); 
 \node at (\ya, -.4 - \vs){$m_B$};
 
 \draw[thick] (\yh,-.1 - \vs)--(\yh,.1 - \vs); 
 \draw[thick, dotted] (\yh, 0 - \vs)--(\yh,1 - \vs);
 \node at (\yh+0.1, -.4 - \vs){$\Lambda_\mathrm{QCD}$};

 \draw[thick] (\yr,-.1 - \vs)--(\yr,.1 - \vs); 
 \node at (\yr, -.4 - \vs){$\mu_0$};

 \draw[thick] (\yb,-.1 - \vs)--(\yb,.1 - \vs); 
 \node at (\yb -0.0 , -.4 - \vs){$m_\tau$};
 
 \draw[thick] (\yd,-.1 - \vs)--(\yd,.1 - \vs); 
 \node at (\yd, -.35 - \vs){$E_{\rm cut}$};

 \node at (\ya*0.5+\y0*0.5,.5 - \vs){\footnotesize{LEFT}};
 \node at (\yh*0.5+\ya*0.5,.5 - \vs){\footnotesize{HFET}}; 
 \node at (\yh*0.5+\ylo*0.5,.5 - \vs){\footnotesize{HMET}};
\end{tikzpicture} 
  
\end{center}
\end{minipage}\vspace{-3ex}

This paper is structured in the following way: In Section \ref{sec:hqet}, we describe the construction of the HFET including first-order $1/m_\tau$ corrections, the derivation of the associated matching coefficients and their renormalization-group (RG) evolution. We then give a brief description of HMET in Section~\ref{sec:hhchipt}, along with the derivation of decay rates and RG equations, leading to the factorization formula for the exclusive $B^-\to\tau^-\spac\bar\nu_\tau(\gamma)$ decay rate. In Section~\ref{sec:pheno} we  present numerical estimates for the rate. Combining the results of the present study with those obtained in \cite{Cornella:2026lkp}, we also derive a state-of-the-art prediction for the lepton-flavor universality (LFU) ratio $R_{\tau\mu}$ between the rates for the tau and muon channels. Our conclusions are given in Section~\ref{sec:conclusions}.

\section{Heavy-Fermion Effective Theory}\label{sec:hqet}

The starting point of our analysis is the LEFT~\cite{Jenkins:2017jig}. The only relevant operator in the SM is
\begin{align}\label{eq:LLEFT}
   \mathcal L_\mathrm{LEFT} 
   = - \frac{4 G_F^{(\mu)}}{\sqrt{2}}\spac V_{ub}\,K_\mathrm{EW}(\mu)\spac 
    \big(\bar u\gamma^\mu P_L\spac b\big) \big(\bar\tau\gamma_\mu P_L\spac\nu_\tau\big) \,,
\end{align}
where $G_F^{(\mu)}$ denotes the Fermi constant extracted from muon decay, $V_{ub}$ is the relevant CKM matrix element, and
\begin{align}
   K_\mathrm{EW}(\mu) 
   = 1 + \frac{Q_\tau\spac\alpha}{4\pi} \left[
    3 Q_u \left( \ln\frac{\mu^2}{m_Z^2} + \frac{11}{6} \right) 
    + (Q_b+Q_u) \left( 1 + \frac{\kappa}{4} \right) \right] + \mathcal{O}(\alpha^2)
\end{align}
encodes the electroweak matching corrections at one-loop order~\cite{Marciano:1993sh,Bigi:2023cbv}. The parameter $\kappa$ describes the scheme dependence of the Dirac reductions used in the matching procedure. It will be convenient for us to work in a scheme with $\kappa=0$, but we stress that this is not the standard choice, so care must be taken when using results from the literature. We refer the reader to Appendix~A of~\cite{Cornella:2026lkp} for a detailed discussion.

There are in principle two choices for how to treat the tau lepton in our EFT construction. Since $m_\tau/m_B\approx 0.337$ is not a particularly small parameter, it seems natural to treat both the tau lepton and the $b$-quark as heavy particles below the scale $\mu\sim m_B$. This offers the advantage that the exact dependence on the mass ratio $r=m_\tau/m_B$ is kept in the matching to HFET. Logarithms of $r$ are then not resummed, but given that $\ln(1/r)\approx 1.09$ such a resummation is not expected to be numerically important. Alternatively, one could adopt the scaling $m_\tau^2\sim\Lambda_{\rm QCD}\spac m_B$ and treat the tau lepton below $\mu\sim m_B$ as a hard-collinear particle in soft-collinear effective theory (SCET), as was done for the muon case in \cite{Cornella:2026lkp}. In such a treatment, the Wilson coefficients in the matching onto SCET are independent of $m_\tau$, and hence terms of the form $\left(m_\tau/m_B\right)^n$ (with $n\ge 1$) are neglected. When hard-collinear particles are integrated out at the scale $\mu_0$, the Wilson coefficients are jet functions depending on the two scales $\Lambda_{\rm QCD}\spac m_B$ and $m_\tau^2$. This resums a series of terms of the form $\left(\Lambda_{\rm QCD}\spac m_B/m_\tau^2\right)^n$, which are not resummed in the alternative approach mentioned above. We will show in this paper that the term with $n=1$ is absent at $\mathcal{O}(\alpha)$. Moreover, analogous studies of $\left(\Lambda_{\rm QCD}\spac m_B/m_c^2\right)^n$ corrections in rare $B$-meson decays involving charm-quark loops \cite{Voloshin:1996gw,Ligeti:1997tc,Grant:1997ec,Buchalla:1997ky,Benzke:2010js} have indicated that the higher-order terms in this series (with $n\ge 2$) are associated with small numerical coefficients, so that their resummation might not be relevant. For these reasons, we treat the tau lepton and the $b$-quark as heavy particles in this paper.

At the hard matching scale $\mu_h\sim m_B\sim m_\tau$, we match the four-fermion operator in \eqref{eq:LLEFT} onto HFET, in which both the $b$ quark and the tau lepton are described by effective heavy-fermion spinor fields $b_v$ and $\tau_{v_\tau}$ defined as \cite{Neubert:1993mb}
\begin{align}
   b_v(x) = e^{im_b v\cdot x}\,\frac{1+\slashed v}{2}\,b(x) \,, \qquad
   \tau_\vl(x) = e^{im_\tau\vl\cdot x}\,\frac{1+\slashed v_\tau}{2}\,\tau(x) \,,
\end{align}
where $v$ and $v_\tau$ (with $v^2=v_\tau^2=1$) are the 4-velocities of the $B$ meson and the tau lepton, respectively. The original fields $b(x)$ and $\tau(x)$ can be expressed as infinite series of the form\footnote{In these relations one often uses the ``spatial'' derivatives $(D^\mu-v^\mu\spac v\cdot D)$ and $(D^\mu-v_\tau^\mu\,v_\tau\cdot D)$, which vanish when contracted with $v_\mu$ and $v_{\tau\spac\mu}$, respectively. However, the second term in these expressions vanishes due to the equations of motion \eqref{eq:EoMs}.} 
\begin{align}\label{eq:field_expansions}
   b(x) = e^{-im_b v\cdot x} 
    \left( 1 + \frac{i\slashed{D}}{2m_b} + \dots \right) b_v(x) \,, \quad
   \tau(x) = e^{-im_\tau v_\tau\cdot x} 
    \left( 1 + \frac{i\slashed{D}}{2m_\tau} + \dots \right) \tau_{v_\tau}(x) \,.
\end{align}
The heavy-particle fields satisfy the constraints $\slashed{v}\spac b_v=b_v$ and $\slashed{v}_\tau\spac\tau_{v_\tau}=\tau_{v_\tau}$, and they obey the equations of motion
\begin{align}\label{eq:EoMs}
   iv\cdot D\,b_v(x) = 0 \,, \qquad
   iv_\tau\cdot D\,\tau_{v_\tau}(x) = 0 \,.
\end{align}
The spectator quark and the neutrino stay unaffected during the hard matching onto HFET, while the momentum modes of the light quark in HFET are restricted to be soft, of order $\Lambda_{\rm QCD}$, and we write the corresponding field as $u_s(x)$.

In this section we are interested in the two-body decay $B^-\to\tau^-\spac\bar\nu_\tau$. In the effective theory above the scale $\Lambda_{\rm QCD}$, additional electromagnetic radiation is kinematically forbidden, since we assume $E_{\rm cut}\ll\Lambda_{\rm QCD}$. In terms of the mass ratio $r=m_\tau/m_B$, the two velocity vectors then obey the relation
\begin{align}
   w \equiv v\cdot\vl = \frac{1+r^2}{2r} \approx 1.654 \,.
\end{align}
In the rest frame of the $B$ meson, the momentum of the neutrino is given by $p_\nu^\mu=E_\nu\spac\barn^\mu$, where $E_\nu$ denotes the neutrino energy and $\barn^\mu$ is a light-like reference vector ($\barn^2=0$) satisfying $\nb\cdot v=1$. The three reference vectors are related to each other using the kinematics of the process, which implies
\begin{align}\label{eq:vtovlnb}
   v^\mu = r\spac\vl^\mu + \frac{1-r^2}{2}\spac\barn^\mu \,.
\end{align}
To construct an EFT basis, it is useful to decompose all 4-vectors into components along $v_\tau$ and $\nb$. For a generic vector $p^\mu$, this leads to the decomposition
\begin{align}\label{eq:vectordecomp}
   p^\mu = \frac{\nb\cdot p}{\nb\cdot v_\tau}\,v_\tau^\mu
    + \left( v_\tau\cdot p - \frac{\nb\cdot p}{\nb\cdot v_\tau} \right) 
    \frac{\nb^\mu}{\nb\cdot v_\tau} + p_\perp^\mu \,,
\end{align}
where the subscript $\perp$ refers to the two Lorentz components orthogonal to the decay plane, and $\nb\cdot v_\tau=m_B/m_\tau$ from \eqref{eq:vtovlnb}.

\subsection{Operator basis at leading power}

After rephasing, the fields $b_v$ and $\tau_\vl$ of the effective theories carry residual momenta of the same order as those of the light quarks, $k^\mu\sim\Lambda_\mathrm{QCD}$, which sets the scale for soft hadronic interactions in HFET. The tau lepton participates in these interactions via soft photon exchange with the quark fields. From this scaling, it follows that the power-counting of the effective theory is done in the parameter $\lambda\sim\Lambda_\mathrm{QCD}/m_B\sim\Lambda_\mathrm{QCD}/m_\tau$. The soft fermion and gauge fields scale as
\begin{align}\label{eq:bvtauvdef}
   b_v, \tau_\vl, u_s \sim \lambda^{3/2} \qquad \text{and} \qquad
   G^\mu, A^\mu \sim \lambda \,,
\end{align}
where $G$ and $A$ refer to the gluon and photon, respectively. Since the neutrino is inert, we do not need to assign a power counting to it. At leading order in the heavy-quark expansion, $\mathcal{O}(\lambda^{9/2})$, the only relevant operators for our process are four-fermion operators of the form $(\bar{u}_s\Gamma_h\spac b_v)(\bar\tau_\vl\Gamma_\ell\,\nu_\tau)$, with $\Gamma_h$ and $\Gamma_\ell$ being arbitrary Dirac structures. 
A basis of such operators can be constructed following the strategy outlined in \cite{Cornella:2026lkp}. Since the neutrino field satisfies $\nbsl\spac\nu_\tau=0$, the only two possible lepton currents are
\begin{align}
   \bar{\tau}_{v_\tau} \gamma_\mu^\perp P_L\spac\nu_\tau \,, \qquad
   \frac{m_\tau}{m_B}\,\bar{\tau}_{v_\tau} P_L\spac\nu_\tau \,.
\end{align}
Here we have used that $\bar{\tau}_{v_\tau}\spac\slashed{v}_\tau=\bar{\tau}_{v_\tau}$, and that currents involving two transverse Dirac matrices can be reduced to the currents above. The tau lepton in the second current is right-handed and hence requires a mass insertion. This explains our choice of the prefactor. Lorentz invariance now implies that the allowed quark currents can be chosen as 
\begin{align}
   \bar u_s\spac\gamma_\perp^\mu P_L\spac b_v \,, \qquad
   \bar u_s\,\nbsl P_L\spac b_v \,, \qquad
   \bar u_s\spac P_R\,b_v \,,
\end{align}
where we have employed that the light quark is massless and hence must be left-handed. The other possible Dirac structures $\slashed{v} P_L$ and $\slashed{v}_\tau P_L$ can be eliminated using relation \eqref{eq:vtovlnb}. We thus obtain the leading-power basis operators\footnote{A variant of $O_0$ in which $\gamma_\perp^\mu P_L$ in the quark current is replaced by $\gamma_\perp^\mu\spac\nbsl\spac P_R$ yields an operator that vanishes by the identities given in (3.24) of \cite{Cornella:2026lkp}.}
\begin{align}\label{eq:LPop}
\begin{aligned}
   O_0 &= \big( \bar u_s\spac\gamma_\perp^\mu P_L\spac b_v \big)
    \big( \bar{\tau}_{v_\tau} \gamma_\mu^\perp P_L\spac\nu_\tau \big) \,, \\[1mm]
   O_1 &= \frac{m_\tau}{m_B}\,\big( \bar u_s\spac\nbsl P_L\spac b_v \big)
    \big( \bar{\tau}_{v_\tau} P_L\spac\nu_\tau \big) \,, \\[-1mm]
   O_2 &= \frac{m_\tau}{m_B}\,\big( \bar u_s\spac P_R\,b_v \big)
    \big( \bar{\tau}_{v_\tau} P_L\spac\nu_\tau \big) \,.
\end{aligned}
\end{align} 
The factorization of the operators into quark and lepton currents is broken when electromagnetic emissions (virtual or real) are taken into account. In this case, the two currents separately are not even gauge invariant. Operators analogous to $O_0$, $O_1$ and $O_2$ also appear in the muon channel, but here they all contribute at leading order in the heavy-fermion expansion. 

The operator $O_0$ does not contribute to $B^-\to\tau^-\spac\bar\nu_\tau$. Indeed, the quark current has a vanishing matrix element with the $B$ meson in the absence of photon emission, which is kinematically forbidden in HFET since $E_{\rm cut}\ll\Lambda_{\rm QCD}$. This operator will, however, contribute to the real-emission $B^-\to\tau^-\spac\bar\nu_\tau\,\gamma$ process, as will be discussed in Section~\ref{sec:hhchipt}.

\subsection{Matching conditions}

\begin{figure}
\centering
\includegraphics[scale=0.5]{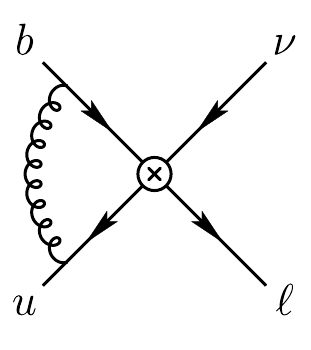}\qquad
\includegraphics[scale=0.5]{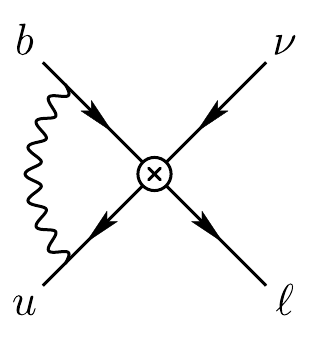}\qquad
\includegraphics[scale=0.5]{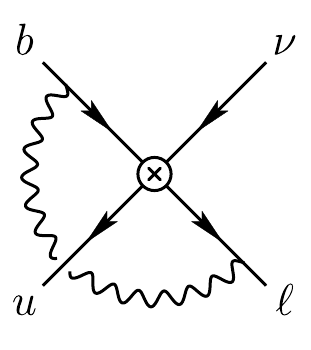}\qquad
\includegraphics[scale=0.5]{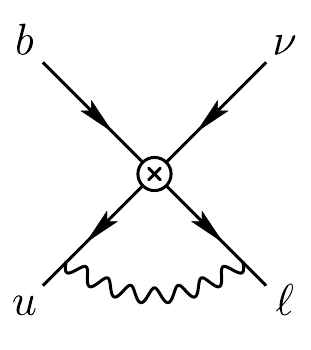}
\caption{\label{fig:hard_match_1loop}
One-loop diagrams contributing to the matching of the LEFT to HFET. Wave-function corrections are not shown.} 
\end{figure}

We write the matching relation from LEFT to HFET in the form
\begin{align}
   \big(\bar u\gamma^\mu P_L\spac b\big) \big(\bar\tau\gamma_\mu P_L\spac\nu_\tau\big)
   = \sum_{i=0,1,2} H_i(\mu)\,O_i(\mu) + \mathcal{O}\bigg(\frac{1}{m_{\tau,B}}\bigg) \,.
\end{align} 
To find the tree-level matching conditions, we insert the expansions of the heavy fields in \eqref{eq:field_expansions} into the four-fermion operator in \eqref{eq:LLEFT}. This yields $H_0=H_1=1$ and $H_2=0$. For the operators $O_1$ and $O_2$ we need to perform the matching at one-loop order. Evaluating the diagrams shown in Figure~\ref{fig:hard_match_1loop} in the $\overline{\rm MS}$ scheme, we obtain the hard functions \begin{align}
\begin{aligned}
   H_1(\mu) 
   &= 1 - \frac{3 C_F\spac\alpha_s}{8\pi} \left( \ln\frac{\mu^2}{m_B^2} + \frac{8}{3} \right) 
    + \frac{\alpha}{2\pi}\,c_1(\mu) \,, \\
   H_2(\mu) &= \frac{C_F\spac\alpha_s}{2\pi} + \frac{\alpha}{2\pi}\,c_2(\mu) \,, 
\end{aligned}
\end{align}
with the one-loop QED coefficients
\begin{align}
\begin{aligned}
   c_1(\mu) &= - Q_\tau^2 \left( \frac{3}{4}\spac\ln\frac{\mu^2}{m_\tau^2} + 1 \right) 
    - Q_b^2 \left( \frac{3}{4}\spac\ln\frac{\mu^2}{m_B^2} + 1 \right) 
    - Q_\tau\spac Q_u \left( \frac32 \ln\frac{\mu^2}{m_\tau^2} + 4 \right)
    - Q_b\spac Q_u \\
   &\quad + Q_\tau\spac Q_b \left[ \ln\frac{\mu^2}{m_B^2} 
    \left( \frac{1+r^2}{1-r^2}\spac\ln\frac{1}{r} + \frac12 \right) 
    + \frac{1+r^2}{1-r^2}\,\ln\frac{1}{r} \left( \ln\frac{1}{r} + 1 \right) \right] , \\
   c_2(\mu) &= Q_b\spac Q_u - \frac{2\spac Q_\tau\spac Q_b}{1-r^2}\spac\ln\frac{1}{r} \,.
\end{aligned}
\end{align}
In deriving these results we used the reductions 
\begin{align}
\begin{aligned}
   \gdirac{\gamma_\perp^\mu\gamma_\perp^\nu P_R}{\gamma_\mu^\perp\gamma_\nu^\perp P_L} 
   &= \frac{4}{1-r^2} \left( \gdirac{P_R}{P_L} - r\spac\gdirac{\slashed v_\tau P_L}{P_L} 
    \right) , \\         
   \gdirac{\slashed{v}_\tau\gamma_\perp^\mu\gamma_\perp^\nu P_L}%
    {\gamma_\mu^\perp\gamma_\nu^\perp P_L} 
   &= \frac{4r}{1-r^2} \left( \gdirac{P_R}{P_L} - r\spac\gdirac{\slashed v_\tau P_L}{P_L} 
    \right) ,
\end{aligned}
\end{align} 
which hold in our reduction scheme with the previously explained choice $\kappa = 0$. The lepton mass appearing in the operator definitions in \eqref{eq:LPop} denotes the physical (pole) mass of the tau lepton. 

To resum large logarithms of the form $\ln(m_B/\mu_0)$, where $\mu_0$ denotes the scale at which the hadronic matrix elements are evaluated, the hard functions are calculated at their natural scale, $\mu_h\sim m_B\sim m_\tau$, and evolved to the scale $\mu_0\ll m_B$ using the RG. The corresponding anomalous dimensions are straightforward to derive, as there is no operator mixing at leading power in HFET. For the first two hard functions, we obtain 
\begin{align}\label{eq:rg_H12}
   \frac{1}{H_{1,2}}\,\frac{d}{d\ln\mu}\,H_{1,2}(\mu) 
   = \gamma_{\rm hl}(\alpha_s) + \frac{\alpha}{\pi}
    \left[ \spac \frac34 \left( Q_\tau^2 - Q_b^2 \right) 
    + Q_\tau\spac Q_b\,\gamma_\mathrm{soft}(w) \right] 
    + \mathcal{O}(\alpha_s\spac\alpha,\alpha^2) \,, 
\end{align}
where $\gamma_\mathrm{soft}(w)$ is the well-known soft anomalous dimension \cite{Korchemsky:1987wg,Falk:1990yz}
\begin{align}\label{eq:gammasoft}
   \gamma_\mathrm{soft}(w) 
   = \frac{w}{\sqrt{w^2-1}}\,\ln\big( w + \sqrt{w^2-1} \big) - 1 
   = \frac{1+r^2}{1-r^2}\spac\ln\frac{1}{r} - 1 \,,
\end{align}
which is positive for $w>0$. The evolution of the hard functions in pure QCD is governed by the universal anomalous dimension of heavy-light current operators in HQET, whose two-loop expression reads \cite{Ji:1991pr}
\begin{align}\label{eq:gammahl}
   \gamma_{\rm hl}(\alpha_s) 
   = - \frac{3 C_F\spac\alpha_s}{4\pi} 
    - \left( \frac{254}{9} + \frac{56\spac\pi^2}{27} - \frac{20}{9}\,n_f \right) 
    \left( \frac{\alpha_s}{4\pi} \right)^2 
    + \mathcal{O}(\alpha_s^3) \,.
\end{align}
Here $n_f=4$ is the appropriate number of active quark flavors below the scale $m_B$, and the two-loop coefficient has been evaluated for $N_c=3$. For completeness, we also quote the one-loop evolution equation for the electroweak matching factor $K_\mathrm{EW}$, which reads
\begin{align}\label{eq:rg_KEW}
   \frac{1}{K_\mathrm{EW}}\,\frac{d}{d\ln\mu}\,K_\mathrm{EW}(\mu) 
   = Q_\tau\spac Q_u\,\frac{3\alpha}{2\pi} + \mathcal{O}(\alpha^2) \,.
\end{align}
The solutions to the evolution equations to leading-logarithmic accuracy in QED are given by 
\begin{align}
   H_{1,2}(\mu_0) = U_{\rm hl}(\mu_0,\mu_h) 
    \left( \frac{\mu_h}{\mu_0} \right)^{-\frac{\alpha}{3\pi} 
    \left[ \gamma_{\rm soft}(w) + 2 \right]} H_{1,2}(\mu_h) \,,
\end{align}
and 
\begin{equation}
   K_\mathrm{EW}(\mu_0) \big|_{\kappa=0}
   = \left( \frac{\alpha(m_Z)}{\alpha(\mu_h)} \right)^{\frac{9}{40}}
    \left( \frac{\mu_h}{\mu_0} \right)^{\frac{\alpha}{\pi}}
    \left( 1 - \frac{\alpha}{\pi} \right) ,
\end{equation}
where we have inserted the physical values of the quark charges. The explicit expression for the QCD evolution function $U_{\rm hl}(\mu_0,m_B)$ at NLO in RG-improved perturbation theory is given in (3.90) of \cite{Cornella:2026lkp}. Following this reference, we neglect the running of the electromagnetic coupling $\alpha(\mu)$ in the region below the hard matching scale $\mu_h$, which is numerically tiny. In our discussion below, we fix the hard matching scale to be $\mu_h=m_B$, but varying it over the region between $m_B$ and $m_\tau$ would have a negligible impact on our numerical results.

\subsection[Operator basis including first-order $1/m_\tau$ corrections]%
{\boldmath Operator basis including first-order $1/m_\tau$ corrections}
\label{subsec:2.3}

Power corrections in HFET can be subdivided into those stemming from either the expansion in $1/m_b$ or $1/m_\tau$, with the leading ones arising from the latter. Here we will provide a full account of the $1/m_\tau$ corrections at tree level, which will be sufficient for our purposes. They consist of corrections to the four-fermion operators as well as time-ordered products of the leading operators $O_0$ and $O_1$ with the $1/m_\tau$ corrections to the HFET Lagrangian. To find the power corrections to the four-fermion operators, we perform the tree-level matching of LEFT onto HFET including the $1/m_\tau$ term in the expansion of the tau-lepton field in \eqref{eq:field_expansions}. Decomposing the encountered Dirac matrices using \eqref{eq:vectordecomp}, we obtain (setting $x=0$ for simplicity)
\begin{align}\label{eq:treematch}
   \big(\bar u\gamma^\mu P_L\spac b\big) \big(\bar\tau\gamma_\mu P_L\spac\nu_\tau\big)
   = O_0 + O_1 + \frac{1}{2m_\tau}\,\big[ O_3 + O_4 + O_5 + O_6 \big]
    + \mathcal{O}\bigg( \frac{1}{m_B} \bigg) \,,
\end{align}
with 
\begin{align}\label{eq:NLP1}
\begin{aligned}
   O_3 &= \big( \bar u_s\spac\gamma_\perp^\mu P_L\spac b_v \big)
    \big( \bar{\tau}_{v_\tau} (i\slashed{D}_\perp)^\dagger\gamma_\mu^\perp P_L\spac\nu_\tau \big)
    = \big( \bar u_s\spac\gamma_\nu^\perp\spac\nsl\spac P_R\,b_v \big)
    \big( \bar{\tau}_{v_\tau} (iD_\perp^\nu)^\dagger P_L\spac\nu_\tau \big) \,, \\[2mm]
   O_4 &= \frac{m_\tau}{m_B}\,\big( \bar u_s\spac\gamma_\perp^\mu P_L\spac b_v \big)
    \big( \bar{\tau}_{v_\tau} (i\nb\cdot D)^\dagger\gamma_\mu^\perp P_L\spac\nu_\tau \big) \,, \\
   O_5 &= - \frac{m_\tau}{m_B}\,\big( \bar u_s\spac\nbsl P_L\spac b_v \big)
    \big( \bar{\tau}_{v_\tau} (i\slashed{D}_\perp)^\dagger P_L\spac\nu_\tau \big) \,, \\[-1mm]
   O_6 &= - \left( \frac{m_\tau}{m_B} \right)^2 \big( \bar u_s\spac\nbsl P_L\spac b_v \big)
    \big( \bar{\tau}_{v_\tau} (i\nb\cdot D)^\dagger P_L\spac\nu_\tau \big) \,.
\end{aligned}
\end{align} 
where $\nsl=2\vsl-\nbsl$, $(iD)^\dagger=(-i\!\spac\overleftarrow{D})$, and we have used the equation of motion \eqref{eq:EoMs} for the tau-lepton field. The form of $O_3$ has been simplified using the identities \cite{Lange:2003pk,Cornella:2026lkp} 
\begin{align}\label{eq:nicerelations}
   \gamma_\nu^\perp\gamma_\mu^\perp P_L\spac\nu_\tau 
    = (g_{\nu\mu}^\perp-i\epsilon_{\nu\mu}^\perp)\spac P_L\spac\nu_\tau \,, \qquad 
   (g_{\nu\mu}^\perp-i\epsilon_{\nu\mu}^\perp)\spac\gamma_\perp^\mu P_L\spac b_v
    = \gamma_\nu^\perp\spac\nsl\spac P_R\,b_v \,,
\end{align}
where $g_{\nu\mu}^\perp$ and $\epsilon_{\nu\mu}^\perp$ are the rank-2 symmetric and antisymmetric tensors in the transverse plane. If desired, one can replace 
\begin{align}
   (i\nb\cdot D)^\dagger \to \frac{2}{1-r^2}\,(iv\cdot D)^\dagger 
\end{align} 
in the operators $O_4$ and $O_6$, using again the equation of motion. 

A second source of $1/m_\tau$ corrections results from the first-order power corrections \cite{Neubert:1993mb}
\begin{align}
   \mathcal{L}_{1/m_\tau}
   = \frac{1}{2m_\tau} \left[ \bar\tau_{\vl} (iD)^2\spac\tau_{\vl}
    + \frac{Q_\tau\spac e}{2}\,\bar\tau_{\vl} \sigma_{\alpha\beta} 
    F^{\alpha\beta} \tau_{\vl} \right] ,
\end{align}
to the tau-lepton sector in the HFET Lagrangian, which can correct the matrix elements of the leading-power operators $O_i$ with $i=0,1,2$. Since the tau-lepton is an on-shell particle, the only contributions relevant at tree level involve a photon field, and they arise only from the second (the ``magnetic'') operator. After a straightforward calculation, we find 
\begin{align}\label{eq:Tproducts}
\begin{aligned}
   T \left\{ O_0,\, i\!\int\!d^4y\,\mathcal{L}_{1/m_\tau}(y) \right\}
   &= \frac{1}{2m_\tau}\,\big[ O_7 + O_8 + O_9 \big] \,, \\
   T \left\{ O_1,\, i\!\int\!d^4y\,\mathcal{L}_{1/m_\tau}(y) \right\}
   &= \frac{1}{2m_\tau}\,\big[ O_{10} + O_{11} + O_{12} \big] \,,
\end{aligned}
\end{align}
with operators   
\begin{align}\label{eq:NLP2a}
\begin{aligned}
   O_7 &= \frac{i\epsilon_{\alpha\beta}^\perp}{2}\,\F_\vl^{\spac\alpha\beta}\,
    \big( \bar u_s\spac\gamma_\perp^\mu P_L\spac b_v \big)
    \big( \bar{\tau}_{v_\tau} \gamma_\mu^\perp P_L\spac\nu_\tau \big) \,, \\
   O_8 &= - v_{\tau\spac\alpha}\spac\F_\vl^{\spac\alpha\beta}\, 
    \big( \bar u_s\spac\gamma_\beta^\perp\spac\nsl\spac P_R\,b_v \big)
    \big( \bar{\tau}_{v_\tau} P_L\spac\nu_\tau \big) \,, \\[1mm]
   O_9 &= \frac{m_\tau}{m_B}\,\nb_\alpha\spac\F_\vl^{\spac\alpha\beta}\, 
    \big( \bar u_s\spac\gamma_\beta^\perp\spac\nsl\spac P_R\,b_v \big)
    \big( \bar{\tau}_{v_\tau} P_L\spac\nu_\tau \big) \,, 
\end{aligned}
\end{align} 
and
\begin{align}\label{eq:NLP2b}
\begin{aligned}
   O_{10} &= - \frac{m_\tau}{m_B}\,
    \frac{i\epsilon_{\alpha\beta}^\perp}{2}\,\F_\vl^{\spac\alpha\beta}\,
    \big( \bar u_s\spac\nbsl P_L\spac b_v \big)
    \big( \bar{\tau}_{v_\tau} P_L\spac\nu_\tau \big) \,, \\
   O_{11} &= - \frac{m_\tau}{m_B}\,v_{\tau\spac\alpha}\spac\F_\vl^{\spac\alpha\beta}
    \big( \bar u_s\spac\nbsl P_L\spac b_v \big)
    \big( \bar{\tau}_{v_\tau} \gamma_\beta^\perp P_L\spac\nu_\tau \big) \,, \\[-1mm]
   O_{12} &= \left( \frac{m_\tau}{m_B} \right)^2 \nb_\alpha\spac\F_\vl^{\spac\alpha\beta}
    \big( \bar u_s\spac\nbsl P_L\spac b_v \big)
    \big( \bar{\tau}_{v_\tau} \gamma_\beta^\perp P_L\spac\nu_\tau \big) \,.
\end{aligned}
\end{align} 
where we have defined 
\begin{align}\label{eq:calFdef}
   \F_\vl^{\spac\alpha\beta}(x)
   \equiv - Q_\tau\spac e \int_0^\infty\!dt\,F^{\alpha\beta}(x+t\spac v_\tau) \,.
\end{align} 
This object arises naturally from the heavy-lepton propagator connecting the two operators in the time-ordered products in \eqref{eq:Tproducts}. The form of the various operators has been simplified using the relations \eqref{eq:nicerelations} as well as
\begin{align}
   i\sigma_{\alpha\beta}^\perp\,\nbsl
   = - i\epsilon_{\alpha\beta}^\perp\spac\gamma_5\,\nbsl \,,
\end{align}
where $\nbsl$ is contained in the neutrino field.

The set of operators $\{O_3,\dots,O_{12}\}$ forms a tree-level basis for the $1/m_\tau$-suppressed contributions to the $B^-\to\tau^-\spac\bar\nu_\tau$ decay amplitude. Note that the operators $O_4$ and $O_7$ have a vanishing projection on the $B$ meson and thus can be dropped. The operators $O_3$, $O_8$ and $O_9$ have non-zero matrix elements with the $B$ meson only if the photon field in the covariant derivative is included in the definition of the hadronic matrix elements. On the contrary, the $B$-meson matrix elements of the operators $O_5$, $O_{11}$ and $O_{12}$ would be non-vanishing only if the photon field is {\em not\/} included in the definition of the hadronic matrix element. The photon would then need to be emitted into the final state, which is kinematically forbidden in HFET above the scale $\Lambda_{\rm QCD}$.

\subsection{Hadronic matrix elements}

The next task is to evaluate the $B^-\to\tau^-\spac\bar\nu_\tau$ matrix elements of the basis operators $O_i$ at the scale $\mu_0$ including virtual QED effects, which connect the lepton currents to the quark currents. It is convenient to perform the soft decoupling transformation \cite{Bauer:2001yt}, which removes all (leading-power) gauge interactions from the effective tau-lepton field. We define 
\begin{align}
   \bar\tau_\vl(x) = \bar\tau_\vl^{(0)}(x)\,Y_{v_\tau}^{(\tau)\dagger}(x) \,,
\end{align}
which introduces the soft Wilson line 
\begin{align}\label{eq:Ytau}
   Y_\vl^{(\tau)\dagger}(x) 
   = \exp\left[ i\spac Q_\tau\spac e \int_0^\infty\!dt\,\vl\cdot A(x+t\spac\vl) \right] .
\end{align}
The redefined field $\bar\tau_\vl^{(0)}$ no longer couples to photons and is on-shell, so that its residual momentum vanishes. The basis operators at leading power then take the form
\begin{align}
\begin{aligned}
   O_0 &= \big( \bar u_s\spac\gamma_\perp^\mu P_L\spac b_v\spac Y_\vl^{(\tau)\dagger} \big)
    \big( \bar{\tau}_{v_\tau}^{(0)} \gamma_\mu^\perp P_L\spac\nu_\tau \big) \,, \\[1mm]
   O_1 & = \frac{m_\tau}{m_B}\,\big( \bar u_s\,\nbsl P_L\spac b_v\spac 
    Y_\vl^{(\tau)\dagger}\big)
    \big( \bar{\tau}_{v_\tau}^{(0)} P_L\spac\nu_\tau \big) \,, \\
   O_2 &= \frac{m_\tau}{m_B}\,\big( \bar u_s\spac P_R\,b_v\spac Y_\vl^{(\tau)\dagger} \big)
    \big( \bar{\tau}_{v_\tau}^{(0)} P_L\spac\nu_\tau \big) \,. 
\end{aligned}
\end{align} 
The matrix elements of the lepton currents are now simply given by on-shell spinor products. Notice that the presence of the leptonic Wilson line renders the hadronic currents gauge invariant. An important difference with regard to the muon case is that here gauge invariance is restored by means of a time-like Wilson line, whereas a light-like Wilson line appeared in the corresponding hadronic currents in (4.18) of \cite{Cornella:2026lkp}, which gave rise to a more singular behavior under renormalization. The presence of a time-like Wilson line makes our hadronic matrix elements likely more easily accessible to an evaluation in lattice field theory.

The subleading-power operators $O_{7,\dots,12}$ transform in a similar way, i.e.\ one replaces $\bar{\tau}_{v_\tau}\to\bar{\tau}_{v_\tau}^{(0)}$ and includes the Wilson line $Y_\vl^{(\tau)\dagger}$ into the quark currents. For the case of the operators $O_{3,\dots,6}$, one encounters in addition the gauge-invariant soft photon field \cite{Bauer:2002nz,Hill:2002vw}
\begin{align}\label{eq:calAdef}
   \A_{\vl}^\mu(x) = Y_\vl^{(\tau)\dagger} (iD^\mu)^\dagger Y_\vl^{(\tau)}
   = Y_\vl^{(\tau)\dagger} iD^\mu\spac Y_\vl^{(\tau)} 
   = - Q_\tau\spac e \int_0^\infty\!dt\,v_{\tau\spac\alpha}\spac 
    F_s^{\alpha\mu}(x+t\spac v_\tau) \,,
\end{align}
which satisfies $v_\tau\cdot\A_\vl=0$. In the second step we have used that the Wilson line is a unitary operator. It then follows that
\begin{align}\label{eq:O3to6final}
\begin{aligned}
   O_3 &= \A_{\vl}^\beta
    \big( \bar u_s\spac\gamma_\beta^\perp\spac\nsl\spac P_R\,b_v\spac 
    Y_\vl^{(\tau)\dagger} \big)
    \big( \bar{\tau}_{v_\tau}^{(0)} P_L\spac\nu_\tau \big) \,, \\[2mm]
   O_4 &= \frac{m_\tau}{m_B}\,\nb\cdot\A_{\vl}
    \big( \bar u_s\spac\gamma_\perp^\mu P_L\spac b_v \spac 
    Y_\vl^{(\tau)\dagger} \big)
    \big( \bar{\tau}_{v_\tau}^{(0)} \gamma_\mu^\perp P_L\spac\nu_\tau \big) \,, \\
   O_5 &= - \frac{m_\tau}{m_B}\,\A_{\vl}^\beta
    \big( \bar u_s\spac\nbsl P_L\spac b_v\spac 
    Y_\vl^{(\tau)\dagger} \big)
    \big( \bar{\tau}_{v_\tau}^{(0)} \gamma_\beta^\perp P_L\spac\nu_\tau \big) \,, \\[-1mm]
   O_6 &= - \left( \frac{m_\tau}{m_B} \right)^2 \nb\cdot\A_{\vl}
    \big( \bar u_s\spac\nbsl P_L\spac b_v\spac 
    Y_\vl^{(\tau)\dagger} \big)
    \big( \bar{\tau}_{v_\tau}^{(0)} P_L\spac\nu_\tau \big) \,.
\end{aligned}
\end{align} 
The definitions \eqref{eq:calFdef} and \eqref{eq:calAdef} imply that $v_{\tau\spac\alpha}\spac\F_\vl^{\alpha\beta}=\A_\vl^\beta$, which allows us to identify the operators
\begin{align}
   O_8 = - O_3 \,, \qquad
   O_{11} = O_5 \,.
\end{align} 
Interestingly, the leading $1/m_\tau$-suppressed operator $O_3$, whose contribution to the decay amplitude would scale like $(\Lambda_{\rm QCD}\spac m_B)/m_\tau^2\approx 1$ relative to the contribution of the operator $O_1$, because it does not come with the chiral factor $m_\tau/m_B$, is exactly cancelled by the time-ordered product contribution $O_8$. Dropping operators with a vanishing projection on the $B$ meson, or which would require a (kinematically forbidden) photon emission to be non-zero, we find that the complete expression for the $1/m_\tau$-suppressed operators at tree level takes the final form
\begin{align}
   \frac{1}{2m_\tau}\,\big( O_6 + O_9 + O_{10} \big) \,.
\end{align} 
When written out explicitly, the result reads
\begin{align}\label{eq:NLPfinal}
\begin{aligned}
   \frac{1}{2m_\tau}\,\big( \bar{\tau}_{v_\tau}^{(0)} P_L\spac\nu_\tau \big)\spac
   &\bigg[ \frac{m_\tau}{m_B}\,\nb_\alpha\spac\F_\vl^{\spac\alpha\beta}\, 
     \big( \bar u_s\spac\gamma_\beta^\perp\spac\nsl P_R\,b_v\spac 
     Y_\vl^{(\tau)\dagger} \big) 
     - \frac{m_\tau}{m_B}\,\frac{i\epsilon_{\alpha\beta}^\perp}{2}\,\F_\vl^{\spac\alpha\beta}\,
     \big( \bar u_s\spac\nbsl P_L\spac b_v\spac Y_\vl^{(\tau)\dagger} \big) \\
   &\hspace{1.55mm} - \left( \frac{m_\tau}{m_B} \right)^2 \nb\cdot\A_{\vl}\spac
     \big( \bar u_s\spac\nbsl P_L\spac b_v\spac Y_\vl^{(\tau)\dagger} \big)
    \bigg] \,.
\end{aligned}
\end{align} 

We use the HQET trace formalism to parameterize the hadronic matrix elements of the gauge-invariant quark currents in terms of universal (i.e.\ heavy-quark-mass independent) non-perturbative parameters. For the leading-order currents, we define \cite{Neubert:1992fk}
\begin{align}\label{eq:F1F2def}
   \frac{\langle\spac 0\spac|\spac\bar u_s\spac\Gamma\,b_v\spac Y_\vl^{(\tau)\dagger}
    |B^-(v)\rangle}{R^{(\tau,B)}}
   = - \frac{i\sqrt{m_B}}{2}\,\text{Tr}\bigg\{ 
    \left[ F_1(w,\mu)  +  \frac{\slashed{v}_\tau}{v\cdot v_\tau}\spac F_2(w,\mu) \right] 
    \Gamma\,\frac{1+\slashed{v}}{2}\spac\gamma_5 \bigg\} \,,
\end{align} 
where $\Gamma$ can be a generic Dirac structure. Without the presence of the leptonic Wilson line, i.e.\ in pure QCD, the matrix element is insensitive to the tau-lepton velocity and only the form factor $F_1$ appears. It follows that $F_2$ must be of $\mathcal{O}(\alpha)$. Note that, in the presence of QED effects, the reduced matrix elements $F_i$ are no longer constants but functions of the velocity-transfer variable $w=v\cdot v_\tau$. Hence, they should be interpreted as QED-induced $B^-\to\tau^-$ ``form factors'' rather than ```decay constants''. Moreover, it has been shown in \cite{Beneke:2019slt,Beneke:2020vnb} that the above matrix elements contain infrared divergences from soft photons, due to the long-range nature of the electromagnetic interaction. These divergences must be eliminated by dividing the hadronic currents by the vacuum matrix element 
\begin{equation}
   R^{(\tau,B)} \equiv \langle\spac 0\spac|\,Y_{v_\tau}^{(\tau)\dagger}\,
    \overline{Y}_v^{(B)}\spac |\spac 0\spac\rangle 
\end{equation}
involving an outgoing soft Wilson line for the tau lepton, see \eqref{eq:Ytau}, and an ingoing soft Wilson line for the $B^-$ meson, defined as
\begin{align}\label{eq:YB}
   \overline{Y}_v^{(B)}(x) 
   = \exp\left[ i\spac Q_B\spac e \int_{-\infty}^0\!dt\,v\cdot A(x+t\spac v) \right] .
\end{align}
Only when this is done, the hadronic matrix elements become independent of unphysical infrared regulators. For the matrix elements of the operators $O_1$ and $O_2$, we now obtain 
\begin{align}\label{eq:Fidef}
\begin{aligned}
   \frac{\langle\spac 0\spac|\spac \bar u_s\spac\nbsl P_L\spac b_v\spac 
    Y_\vl^{(\tau)\dagger}\spac|B^-(v)\rangle}{R^{(\tau,B)}}
   &= - \frac{i\sqrt{m_B}}{2} \left[ F_1(w,\mu) 
    - \frac{\nb\cdot v_\tau}{w}\spac F_2(w,\mu) \right] , \\
   \frac{\langle\spac 0\spac|\spac \bar u_s\spac P_R\,b_v\spac 
    Y_\vl^{(\tau)\dagger}\spac|B^-(v)\rangle}{R^{(\tau,B)}}
   &= - \frac{i\sqrt{m_B}}{2}\,\big[ F_1(w,\mu) - F_2(w,\mu) \big] \,,
\end{aligned}
\end{align} 
where $(\nb\cdot v_\tau)/w=2/(1+r^2)$. We parameterize the structure-dependent QED corrections encoded in these parameters in the form
\begin{align}
\begin{aligned}
   F_1(w,\mu) - \frac{\nb\cdot v_\tau}{w}\spac F_2(w,\mu) 
   &\equiv F_{\rm QCD}(\mu) \left[ 1 + \frac{\alpha}{\pi}\,f_1(w,\mu) \right] , \\
   F_1(w,\mu) - F_2(w,\mu) 
   &\equiv F_{\rm QCD}(\mu) \left[ 1 + \frac{\alpha}{\pi}\,f_2(w,\mu) \right] ,
\end{aligned}
\end{align}
where $F_{\rm QCD}$ is the heavy-meson decay constant in HQET without QED effects. The $B$-meson decay constant in pure QCD, $f_B$, is given by the combination
\begin{align}
   \sqrt{m_B}\spac f_B 
   = \left[ H_1(\mu_0) + H_2(\mu_0) \right]_{\rm QCD} F_{\rm QCD}(\mu_0)
    + \mathcal{O}\bigg(\frac{\Lambda_{\rm QCD}}{m_b}\bigg) \,,
\end{align} 
which is scale invariant. Here $[H_i(\mu_0)]_{\rm QCD}$ denotes the values of the hard functions in the absence of QED corrections. The trace formalism can also be used to parameterize the matrix elements of the subleading-power operators in \eqref{eq:NLPfinal}, but given that we only need a single combination of such operators, we simply define
\begin{align}
\begin{aligned}
   & \frac{\langle\spac 0\spac|\spac 
    \nb_\alpha\spac\F_\vl^{\spac\alpha\beta}\, 
     \bar u_s\spac\gamma_\beta^\perp\spac\nsl P_R\,b_v\spac 
     Y_\vl^{(\tau)\dagger} 
     - \big( \frac{i\epsilon_{\alpha\beta}^\perp}{2}\,\F_\vl^{\spac\alpha\beta}
     + \frac{m_\tau}{m_B}\,\nb\cdot\A_{\vl} \big)\spac
     \bar u_s\spac\nbsl P_L\spac b_v\spac Y_\vl^{(\tau)\dagger}   
     \spac|B^-(v)\rangle}{R^{(\tau,B)}} \\
   &\equiv - \frac{i\sqrt{m_B}}{2}\,\Lambda_c\spac F_{\rm QCD}(\mu)\,
    \frac{\alpha}{\pi}\,f_3(w,\mu) \,,
\end{aligned}
\end{align} 
where $\Lambda_c\equiv 500$\,MeV is a typical hadronic scale. Due to the presence of the photon field in the operators, this matrix element would vanish in the limit where QED effects are neglected. The dimensionless functions $f_i(w,\mu)$ depend on the variable $w=v\cdot v_\tau$ and are expected to be of order unity.

For the RG equations for the quantities $F_{1,2}(w,\mu)$ we obtain by explicit calculation 
\begin{align}
   \frac{1}{F_{1,2}}\,\frac{d}{d\ln\mu}\,F_{1,2}(w,\mu)
   = \frac{3 C_F\spac\alpha_s}{4\pi} + \frac{\alpha}{4\pi}\, 
    \Big[ 3\spac Q_u^2 - 4\spac Q_\tau\spac Q_u\spac\gamma_\mathrm{soft}(w) \Big] 
    + \mathcal{O}(\alpha\spac\alpha_s, \alpha^2) \,, 
\end{align} 
which implies that at one-loop order 
\begin{align}\label{eq:f12mudep}
   \frac{d}{d\ln\mu}\,f_{1,2}(w,\mu)
   = \frac34\,Q_u^2 - Q_\tau\spac Q_u\spac\gamma_\mathrm{soft}(w) \,. 
\end{align} 

\subsection{QED-corrected decay amplitude}

We are now ready to compute the $B^-\to\tau^-\spac\bar\nu_\tau$ decay amplitude including virtual QED corrections. We choose to evaluate all component functions at a hadronic (yet still perturbative) matching scale $\mu_0=1.5\,\mathrm{GeV}$, where they are free from large logarithmic corrections. We parameterize the amplitude in the form
\begin{align}\label{eq:hqet_matrix2pelement}
   \mathcal{M}(B^-\to\tau^-\bar\nu_\tau)_{\mu_0} 
   = i\sqrt{2}\,G_F^{(\mu)}\spac V_{ub}\,m_\tau\spac f_B\,
    \bar{u}(p_\tau) P_L\spac v(p_\nu)\,T(\mu_0) \,,
\end{align}
where to first order in $\alpha$, but including the resummation of large logarithmic corrections, we obtain 
\begin{align}\label{eq:Tmu0}
   T(\mu_0) 
   &= K_{\rm EW}(\mu_0) \left\{ 
    \left[ H_1(\mu_0) \right]_{\rm QED} \left[ 1 + \frac{\alpha}{\pi}\,f_1(w,\mu_0) \right] 
    + \left[ H_2(\mu_0) \right]_{\rm QED}
    + \frac{\Lambda_c}{2m_\tau}\,\frac{\alpha}{\pi}\,f_3(w,\mu_0) \right\} \nonumber \\
   &\equiv \left( \frac{\alpha(m_Z)}{\alpha(m_B)} \right)^{\frac{9}{40}}
    \left( \frac{m_B}{\mu_0} \right)^{-\frac{\alpha}{\pi}\spac\gamma_{\rm soft}(w)} 
    \mathcal{R}_{\rm virt} \,,
\end{align}
where $[H_i(\mu_0)]_{\rm QED}$ refer to the hard functions without QCD corrections, and 
\begin{align}
   \mathcal{R}_{\rm virt} 
   = \left( \frac{m_B}{\mu_0} \right)^{\!\frac{\alpha}{3\pi} 
    \left[ 1 + 2\gamma_{\rm soft}(w) \right]} 
    \Bigg\{ 1 + \frac{\alpha}{\pi}\,\bigg[ 
    \underbrace{\frac{1+r^2}{1-r^2}\,\frac{\ln^2 r}{6} - \frac{\ln r}{12} 
     - \frac29}_{\approx\,0.117} 
    + f_1(w,\mu_0) 
    + \underbrace{\frac{\Lambda_c}{2m_\tau}}_{\approx\,0.141}\!f_3(w,\mu_0) \bigg] \Bigg\} \,.
\end{align}
It follows from \eqref{eq:f12mudep} that this quantity is independent of the scale $\mu_0$ to the order we are working. Note that the numerical size of the $r$-dependent term is rather small, despite the fact that it contains a double and a single logarithm of the mass ratio $r$. This supports our treatment, where such logarithms are not resummed. The hadronic parameters $f_i$ evaluated at the low scale $\mu_0$ are free of large logarithmic corrections.  Due to the power suppression of the term proportional to $f_3$, the leading non-perturbative effect is captured by $f_1$. The parameter $f_2$ would enter the above result starting at $\mathcal{O}(\alpha^2)$.

\section{Heavy-Meson Effective Theory}
\label{sec:hhchipt}

Below the hadronic scale, the HFET is matched onto heavy-meson effective theory (HMET), an effective theory of point-like mesons. This matching is genuinely non-perturbative. It has been discussed in detail in Section~5 of \cite{Cornella:2026lkp}. Compared with the muon channel, the present case is structurally simpler, because the tau is not a highly boosted particle in the $B$-meson rest frame. Consequently, there are no ultrasoft-collinear modes in the low-energy theory. Instead, the soft modes of HFET are matched directly onto ultrasoft modes, whose characteristic scale is set by the radiation veto $E_\mathrm{cut}\ll\Lambda_{\rm QCD}$. We define the associated expansion parameter \begin{align}
   \zeta = \frac{E_\mathrm{cut}}{\Lambda_\mathrm{QCD}} \ll 1 \,.
\end{align}
The resulting low-energy theory thus describes point-like mesons and a heavy lepton interacting with ultrasoft photons. Since its construction closely parallels that of the muon channel, we only quote here the ingredients needed for our analysis and refer the reader to \cite{Cornella:2026lkp} for further details. In particular, we forego any discussion of the role of pions in the low-energy theory. 

\begin{figure}
\centering
\includegraphics[scale=0.5]{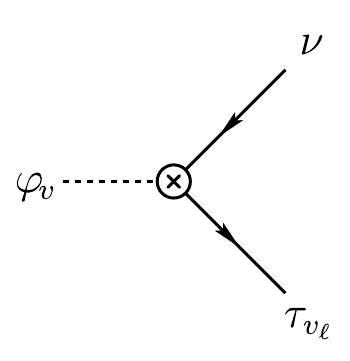}\qquad
\includegraphics[scale=0.5]{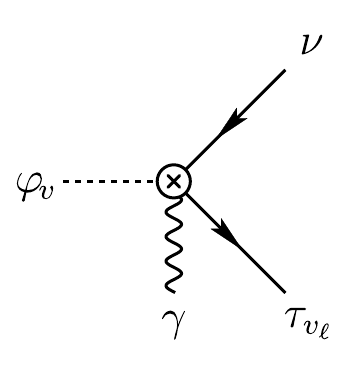}\qquad
\includegraphics[scale=0.5]{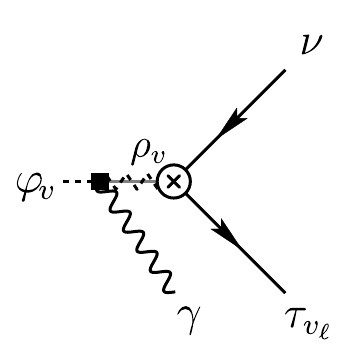}\qquad
\includegraphics[scale=0.5]{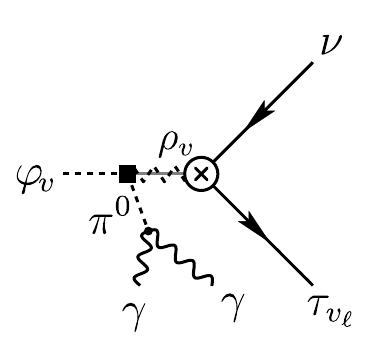}
\caption{\label{fig:real_emission_graphs} 
Feynman diagrams for the decay $B^-\to\tau^-\bar\nu_\tau(\gamma)$ in the effective theory below the hadronic scale. The second graph includes contributions from both the leading-power operators $O_{1,2}$ and the $1/m_\tau$-suppressed operator $O_6$ (see text for explanation). The last two graphs describe the $B$--$B^\ast$ transitions through a photon or a pion, followed by a leptonic decay of the virtual $B^\ast$ meson. The black square indicates that these transitions are power suppressed in $\zeta$.}
\end{figure}

The relevant topologies generated by the HMET are shown in Figure~\ref{fig:real_emission_graphs}. We will refer to the first two graphs as the ``direct'' contributions to the $B^-\to\tau^-\bar\nu_\tau(\gamma)$ decay rate and reserve the term ``indirect'' contributions for decay topologies mediated by an intermediate $B^\ast$ meson decaying to $\tau^-\spac\bar\nu_\tau$. 

\subsection{Direct real-emission contributions}

As in the muon channel, at leading power in $\zeta$ all virtual corrections are encoded in the matching coefficients of the low-energy theory, with the dynamics below the hadronic scale generating only real-emission contributions. The leading terms in the weak-interaction Lagrangian in HMET read \cite{Cornella:2026lkp}
\begin{align}\label{eq:HMETLag}
\begin{aligned}
   \mathcal{L}_{H\ell}
   &= i\sqrt2\spac G_F^{(\mu)}\spac V_{ub}\,Y_{v_\tau}^{(\tau)\dagger}\,\overline{Y}_v^{(B)} \\
   &\quad \times \left\{
    y_B\spac\frac{f_B\spac m_\tau}{\sqrt{2\spac m_B}}\,
    \varphi_v^{(0)}\,\bar\tau_{v_\tau}^{(0)}\spac P_L\spac\nu_\tau
    - \frac{f_{B^\ast}\spac m_{B^\ast}}{\sqrt{2\spac m_{B^\ast}}}\,
    \bigg[ y_{B^\ast}^\perp\spac
    \bar\tau_{v_\tau}^{(0)}\spac\rlap{\spac/}{\rho}_v^{\perp(0)} P_L\spac\nu_\tau
    + y_{B^\ast}^\parallel\spac
    \frac{\nb\cdot\rho_v^{(0)}}{\nb\cdot v_\tau}\,\bar\tau_{v_\tau}^{(0)} P_L\spac\nu_\tau \bigg]
    \right\} ,
\end{aligned}
\end{align}
where $\varphi_v^{(0)}$ and $\rho_v^{(0)}$ are the effective $B$ and $B^\ast$ meson fields in the heavy-hadron limit. All fields (apart from the inert neutrino) obey ultrasoft scaling. This includes in particular the Wilson lines, for which we use the same symbols as for the soft Wilson lines out of notational convenience. The factors $\sqrt{2\spac m_{B^{(*)}}}$ in the denominator account for the relativistic normalization of the meson states and disappear when one takes hadronic matrix elements. Note that the meson fields and the field for the tau lepton are decoupled, as indicated by the superscripts, meaning that all ultrasoft interactions are contained in the Wilson lines. Since the Wilson lines are the same as in the definition of the hadronic matrix elements in \eqref{eq:Fidef}, these matrix elements are the non-perturbative matching coefficients needed to map HFET onto HMET \cite{Cornella:2026lkp}, and in the above result we have expressed the parameters $F_i$ in terms of the (QCD-only) decay constants $f_B$ and $f_{B^\ast}$ as well as matching coefficients $y_B$ and $y_{B^\ast}^{\perp,\parallel}$ accounting for the structure-dependent virtual QED corrections. These ``Yukawa couplings'' contain all information about the QED dynamics at and above the hadronic scale. Matching to the amplitude in \eqref{eq:hqet_matrix2pelement} and the corresponding amplitude for the $B^{*-}$ meson gives
\begin{align}\label{eq:yuk_matching}
   y_B(\mu_0) = T(\mu_0) \,, \qquad 
   y_{B^\ast}^\perp = y_{B^\ast}^\parallel = 1 + \mathcal{O}(\alpha) \,,
\end{align}
where the explicit form of $T(\mu_0)$ has been given in \eqref{eq:Tmu0}, and for the $B^\ast$ contributions the tree-level matching relations are sufficient. The two terms in \eqref{eq:HMETLag} are responsible for the leptonic decay of the $B$ and the $B^\ast$, respectively. Their structure is analogous to that encountered in the muon case, with one important difference: the longitudinal coupling of the $B^\ast$ to leptons, $y_{B^\ast}^\parallel$, is not significantly suppressed, since the chiral factor $1/(\nb\cdot v_\tau)=m_\tau/m_B$ is of $\mathcal{O}(1)$ in the tau case.

We now turn to the prediction for the direct contributions to the decay rate at leading power in $\zeta$, but including the emission of ultrasoft photons. It can be written in the form
\begin{align}\label{eq:gammadir}
    \frac{\Gamma_\mathrm{dir}(E_{\rm cut})}{\Gamma_{\rm tree}} 
    = \big[ y_B(\mu_0) \big]^2\spac R(E_\mathrm{cut},\mu_0) \,, 
\end{align}
where 
\begin{equation}\label{eq:treelevelrate}
   \Gamma_\mathrm{tree}
   = \frac{m_\tau^2\,m_B}{8\pi}
    \left( G_F^{(\mu)}\spac|V_{ub}|\spac f_B \right)^2
    \left( 1 - \frac{m_\tau^2}{m_B^2} \right)^2
\end{equation}
denotes the decay rate in the absence of QED corrections. The virtual QED corrections are contained in $y_B(\mu_0)$. At leading power in the low-energy theory, real emissions arise entirely from photon matrix elements of the Wilson lines contained in \eqref{eq:HMETLag}. Imposing that the real photons have total energy smaller than the radiation veto $E_\mathrm{cut}$, we obtain the radiation function
\begin{align}
   R(E_\mathrm{cut},\mu_0) 
   =  \sum_{n=0}^\infty\,\prod_{i=1}^n\spac
    \int\!\frac{d^{d-1}\boldsymbol{q}_i}{(2\pi)^{d-1}\spac 2 E_i}
    \left| \bra{\gamma_{us}^n} Y_\vl^{(\tau)\dagger}\spac\overline{Y}_v^{(B)}
    \ket{0} \right|^2 \theta\Big(E_\mathrm{cut} - \sum_{j=1}^n E_j \Big) \,,
\end{align}
where $\bra{\gamma_{us}^n}$ denotes an out state with $n$ ultrasoft photons with momenta $q_i$ and energies $E_i$ in the $B$-meson rest frame. Evaluating the one-photon contribution to this expression in the $\overline{\rm MS}$ scheme, we obtain 
\begin{align}
\begin{aligned}
    R(E_\mathrm{cut},\mu_0) 
    &= 1 + \frac{Q_\tau^2\spac\alpha}{\pi}\,\bigg\{\!
     \left( 1 + \frac{1+r^2}{1-r^2}\spac\ln r \right) \ln\frac{\mu_0^2}{4E_\mathrm{cut}^2} \\
    &\hspace{2.5cm} + 1 - \frac{1+r^2}{1-r^2}\,\Big[ \ln r + \ln^2 r + \text{Li}_2(1-r^2) \Big] 
     \bigg\} + \mathcal{O}(\alpha^2) \,.
\end{aligned}
\end{align}
Note that the factor in front of the logarithm of $\mu_0^2/(4E_\mathrm{cut}^2)$ equals $-\gamma_{\rm soft}(w)$ in \eqref{eq:gammasoft}, so that at one-loop order the evolution equation for the radiation function reads
\begin{align}\label{eq:rg_Rsoft}
    \frac{1}{R}\,\frac{dR(E_\mathrm{cut},\mu_0)}{d\ln\mu_0} 
    = - \frac{2\spac Q_\tau^2\spac\alpha}{\pi}\,\gamma_\mathrm{soft}(w) \,,
\end{align}
whose solution takes the form (setting $Q_\tau=-1$)
\begin{align}
   R(E_\mathrm{cut},\mu_0) 
   = \left( \frac{\mu_0}{2E_\mathrm{cut}} 
    \right)^{-\frac{2\alpha}{\pi}\spac\gamma_\mathrm{soft}(w)}\,
    \Bigg\{ 1 + \frac{\alpha}{\pi}\,\bigg[ 
    \underbrace{1 - \frac{1+r^2}{1-r^2}\left(\ln r + \ln^2 r + \text{Li}_2(1-r^2) 
    \right)}_{\approx\,-0.712} \bigg] \Bigg\} \,.
\end{align}
The scale dependence of this quantity cancels the scale dependence of $[y_B(\mu_0)]^2$ in the expression \eqref{eq:gammadir} for the direct contribution to the decay rate. 

Like the leading-order operators $O_{0,1,2}$, also the $1/m_\tau$-suppressed operators $O_{3,\dots,12}$ discussed in Section~\ref{subsec:2.3} can be matched to HMET. In the process, the soft photon fields are matched onto ultrasoft fields (for which we will use the same notation for simplicity) that can be emitted into the final state, with energies of order $E_{\rm cut}$. The interference of the corresponding power-suppressed amplitudes with the leading-order emission amplitude gives rise to first-order power corrections to the radiation function $R(E_{\rm cut},\mu_0)$. Operators with insertions of a transverse Dirac matrix in the lepton current ($O_{4,5,7,11,12}$) or the quark current ($O_{3,4,7,8,9}$) as well as operators involving only transverse components of the photon field ($O_{3,5,7,10}$) do not interfere with the leading-order amplitude. The only non-zero contribution thus arises from the operator $O_6$ in \eqref{eq:O3to6final}, which contains two powers of the chiral factor $m_\tau/m_B$. It is matched onto the operator
\begin{align}
   \mathcal{L}_{H\ell}
   = i\sqrt2\spac G_F^{(\mu)}\spac V_{ub}\,Y_{v_\tau}^{(\tau)\dagger}\,\overline{Y}_v^{(B)} 
    \frac{1}{2m_\tau} \left[ - \frac{m_\tau}{m_B}\,\nb\cdot\A_{\vl}\spac
    \frac{f_B\spac m_\tau}{\sqrt{2\spac m_B}}\,
    \varphi_v^{(0)}\,\bar\tau_{v_\tau}^{(0)}\spac P_L\spac\nu_\tau \right] .
\end{align}
After a straightforward calculation, we obtain 
\begin{align}
    \delta R(E_\mathrm{cut},\mu_0)
    = \frac{E_{\rm cut}}{m_\tau}\,\frac{\alpha}{\pi}\,
     \frac{4r}{1-r^2} \left( 1 + \frac{1+r^2}{1-r^2}\spac\ln r \right) .
\end{align}
The final result for the QED correction factor can then be written in the form
\begin{align}\label{eq:totalQED}
\begin{aligned}
    \frac{\Gamma_\mathrm{dir}(E_{\rm cut})}{\Gamma_{\rm tree}} 
    &= \left( \frac{\alpha(m_Z)}{\alpha(m_B)} \right)^{\frac{9}{20}}
     \left( \frac{2E_\mathrm{cut}}{m_B} \right)^{\frac{2\alpha}{\pi}\spac\gamma_{\rm soft}(w)} 
     \\[1mm]
    &\quad\times \Bigg\{ 1 + \frac{\alpha}{\pi}\,\bigg[ 
     \underbrace{\frac59 - \frac{\ln r}{6} 
     - \frac{1+r^2}{1-r^2}\,\Big[ \ln r + \frac23\spac\ln^2 r + \dilog{1-r^2} 
     \Big]}_{\approx\,-0.478} \\[-1mm]
    &\hspace{2.6cm} + \frac{E_{\rm cut}}{m_\tau}\,
     \underbrace{\frac{4r}{1-r^2} \left( 1 + \frac{1+r^2}{1-r^2}\spac\ln r 
     \right)}_{\approx\,-0.557} \bigg] \Bigg\} \\[-1mm]
    &\quad\times \left( \frac{m_B}{\mu_0} \right)^{\!\frac{2\alpha}{3\pi} 
     \left[ 1 + 2\gamma_{\rm soft}(w) \right]} 
    \Bigg\{ 1 + \frac{\alpha}{\pi}\,\bigg[ 2 f_1(w,\mu_0) 
    + \frac{\Lambda_c}{m_\tau}\,f_3(w,\mu_0) \bigg] \Bigg\} \,,
\end{aligned}
\end{align}
where the remaining scale dependence in the last line cancels between the prefactor and the hadronic matrix element $f_1$. This is our final result for the direct contributions, including $1/m_\tau$-suppressed corrections at tree level. The leading dependence on the radiation veto energy is governed by $(E_\mathrm{cut})^{0.234\spac\alpha}$, which is much milder than for the muon case, since the numerical value of $\gamma_{\rm soft}(w)$ is smaller for the tau channel. 

\subsection{Indirect real-emission contributions}

Another source of power corrections to the $B^-\to\tau^-\spac\bar\nu_\tau\gamma$ decay rate results from higher-order terms in the HMET Lagrangian that scale like $\zeta=E_{\rm cut}/\Lambda_{\rm QCD}$. The leading such terms involve the couplings of a $B$ meson to a $B^\ast$ meson and a photon or pion, given by \cite{Cornella:2026lkp}
\begin{align}\label{eq:BBastLag}
   \mathcal L_{BB^\ast} 
   = \frac{e\spac g_{BB^\ast\gamma}}{2}\,
    \Big( v^\mu\rho_v^{\dagger\nu(0)}\spac\varphi_v^{(0)} \tilde F_{\mu\nu} + \mathrm{h.c.} \Big)
    - \frac{g_{BB^\ast\pi}}{\sqrt{2}\spac f_\pi}\,
    \Big( \rho_v^{\dagger\mu(0)}\spac\varphi_v^{(0)}\spac\partial_\mu \pi^0 + \mathrm{h.c.} \Big) 
    \,,
\end{align}
where $\tilde F_{\mu\nu}=\frac12\epsilon_{\mu\nu\alpha\beta}\spac F^{\alpha \beta}$ is the dual field-strength tensor. These terms generate the vertices marked by black squares in the last two diagrams of Figure~\ref{fig:real_emission_graphs}. The corresponding amplitudes do not interfere with the leading-power decay amplitude and hence generate contributions to the decay rate of order $\zeta^2$. The contribution to the $B^-\to\tau^-\spac\bar\nu_\tau\gamma$ rate is
\begin{equation}\label{eq:photonrate}
   \frac{\Gamma_{B^\ast\gamma}(E_{\rm cut})}{\Gamma_{\rm tree}}
   = \frac{m_B^2}{m_\tau^2}\,\frac{f_{B^\ast}^2\spac m_{B^\ast}}{f_B^2\,m_B}\,
    \frac{\alpha}{6\pi} \left( g_{BB^\ast\gamma}\spac E_{\rm cut} \right)^2 
    \left[ (y_{B^\ast}^\perp)^2 + \frac{r^2}{2}\,(y_{B^\ast}^\parallel)^2 \right]
    I\bigg(0,\frac{\delta_{B^\ast}}{E_{\rm cut}}\bigg) \,,
\end{equation}
with $\delta_{B^\ast}=(m_{B^\ast}^2-m_B^2)/(2m_B)\approx 45.4$\,MeV determined by the $B$--$B^\ast$ mass splitting, and $y_{B^\ast}^\perp=y_{B^\ast}^\parallel=1$ at leading order. The power suppression results from the factor $(g_{BB^\ast\gamma}\spac E_{\rm cut})^2$, since $g_{BB^\ast\gamma}\sim1/\Lambda_{\rm QCD}$. Additional dependence on the veto energy arises from the phase-space function 
\begin{align}
   I(0,z) = 3 - 6z - \frac{2}{1+z} + 6 z^2\ln\frac{1+z}{z} \,.
\end{align}
It approaches~1 for $E_{\rm cut}\gg\delta_{B^\ast}$, while for $E_{\rm cut}\lesssim\delta_{B^\ast}$ it is given by a power series in $E_{\rm cut}/\delta_{B^\ast}$ starting with $\frac12(E_{\rm cut}/\delta_{B^\ast})^2-\frac45(E_{\rm cut}/\delta_{B^\ast})^3\pm \dots$, so that in the limit $E_{\rm cut}\to 0$, the indirect rate scales like $E_{\mathrm{cut}}^4$. 

Expression \eqref{eq:photonrate} is structurally identical to that for the muon channel derived in \cite{Cornella:2026lkp}, except that there we dropped the contribution from the longitudinal polarization state of the $B^\ast$ meson owing to its strong chiral suppression. In the muon channel, the prefactor $m_B^2/m^2_\mu$ provides a strong enhancement of the $B^\ast$ contribution, which makes it phenomenologically relevant, especially for larger values of the radiation veto. In the tau channel this enhancement factor is much smaller, rendering the contribution $\Gamma_{B^\ast\gamma}$ numerically negligible.

The pion-induced contribution $B^-\to B^{*-}\spac(\to\tau^-\spac\bar\nu_\tau)\,\pi^0\spac(\to\gamma\gamma)$ is kinematically allowed only if $E_\mathrm{cut}>m_{\pi^0}$. It can be calculated in complete analogy with the muon channel. We find
\begin{equation}\label{eq:pionrate}
   \frac{\Gamma_{B^\ast\pi}(E_{\rm cut})}{\Gamma_{\rm tree}}
   = \frac{m_B^2}{m_\tau^2}\,\frac{f_{B^\ast}^2\spac m_{B^\ast}}{f_B^2\,m_B}\,
    \frac{g_{BB^\ast\pi}^2}{24\pi^2} \left( \frac{E_{\rm cut}}{f_\pi} \right)^2 
    \left[ (y_{B^\ast}^\perp)^2 + \frac{r^2}{2}\,(y_{B^\ast}^\parallel)^2 \right]
    I\bigg(\frac{m_\pi}{E_{\rm cut}},\frac{\delta_{B^\ast}}{E_{\rm cut}}\bigg) \,,
\end{equation}
where the phase-space function $I(z,y)$ is given in (5.119) of \cite{Cornella:2026lkp}. Since the neutral pion decays to electromagnetic radiation with unit branching ratio, the decay $\pi^0\to\gamma\gamma$ does not lead to a QED suppression, so that this channel rapidly becomes the dominant indirect contribution as $E_\mathrm{cut}$ increases beyond the pion mass. However, once again the impact of this indirect contribution is much larger in the muon case, due to the prefactor $m_B^2/m_\mu^2$ vs.\ $m_B^2/m_\tau^2$.

\section{Phenomenology}\label{sec:pheno}

We now present our numerical results for the $B^-\to\tau^-\spac\bar\nu_\tau(\gamma)$ decay rate with a cut $E_{\rm cut}$ on final-state electromagnetic radiation. This definition is chosen by analogy with the muon channel, allowing for an easy comparison between the two channels. The total decay rate can be decomposed as
\begin{align}
   \Gamma\big(B^-\to\tau^-\spac\bar\nu_\tau(\gamma)\big) 
   = \Gamma_\mathrm{dir}(E_{\rm cut}) 
    + \Gamma_{B^\ast\gamma}(E_{\rm cut}) + \Gamma_{B^\ast\pi}(E_{\rm cut}) \,,
\end{align}
where the various contributions have been discussed above and given in \eqref{eq:totalQED}, \eqref{eq:photonrate} and \eqref{eq:pionrate}, respectively. We now give numerical estimates for each of them. In our analysis, all particle masses, the Fermi constant $G_F^{(\mu)}$ and the reference values for the electromagnetic coupling are taken from the Particle Data Group Review~\cite{ParticleDataGroup:2024cfk}. We perform the RG evolution of the electromagnetic coupling between the scales $\mu_Z=m_Z$ and $\mu_h=m_B$ at one-loop order, finding $\alpha(m_B)^{-1}=131.961$. Hadronic input parameters are obtained from a combination of lattice QCD calculations, QCD sum rules, and quark model estimates. Table~\ref{tab:inputs} shows a compilation of the input parameters.

We begin with the direct contribution to the decay rate in \eqref{eq:totalQED} and show the numerical size of its various components. The largest QED correction comes from the electroweak resummation factor
\begin{align}
   \left( \frac{\alpha(m_Z)}{\alpha(m_B)} \right)^{\frac{9}{20}} 
   = 1 + 1.41\cdot 10^{-2} \,. 
\end{align}
Large logarithms between the hard scale $m_B$ and the factorization scale $\mu_0=1.5$\,GeV are resummed into the factor 
\begin{align}
   \left( \frac{m_B}{\mu_0} \right)^{\!\frac{2\alpha}{3\pi} 
    \left[ 1 + 2\gamma_{\rm soft}(w) \right]}
   = 1 + 0.35\cdot 10^{-2} \,.
\end{align}
The resummed logarithmic dependence on the radiation veto turns out to be mild. For three representative values of $E_\mathrm{cut}$, we find
\begin{align}
   \left( \frac{2E_\mathrm{cut}}{m_B} \right)^{\!\frac{2\alpha}{\pi}\spac\gamma_\mathrm{soft}(w)}
   = 1 - 10^{-2}\cdot \left\{ 
    \begin{array}{lll}
     0.86 \,; &\quad E_\mathrm{cut} &= 20~\mathrm{MeV}, \\
     0.70 \,; &\quad E_\mathrm{cut} &= 50~\mathrm{MeV}, \\
     0.58 \,; &\quad E_\mathrm{cut} &= 100~\mathrm{MeV},\\
     0.51 \,; &\quad E_\mathrm{cut} &= 150~\mathrm{MeV}.
    \end{array}
   \right.
\end{align}
The remaining fixed-order QED corrections evaluate to 
\begin{align}
   1 + \bigg[ - 1.15 - 0.076\,\frac{E_{\rm cut}}{100\,\text{MeV}} 
    + \underbrace{4.82\spac f_1(w,\mu_0) + 0.68\spac f_3(w,\mu_0)}_{\pm 4.87}
    \bigg] \cdot 10^{-3} \,,
\end{align}
where the QED-induced $B^-\to\tau^-$ form factors $f_i(w,\mu_0)$ are currently unknown. At the low scale $\mu_0$ they are free of large logarithms, and we will vary their values independently in the range $[-1,+1]$, which yields the indicated uncertainty. We stress that it should be possible to compute these form factors using lattice field theory, as their definition involves only time-like (not light-like) Wilson lines. Also note that the contribution proportional to the veto energy, which originates from the $1/m_\tau$-suppressed operators in the low-energy EFT, remains negligible even for larger values of $E_{\rm cut}$. Combining the above results, we find
\begin{align}
   \frac{\Gamma_\mathrm{dir}}{\Gamma_\mathrm{tree}}
   = 1 + 10^{-2}\cdot \left\{ 
    \begin{array}{lll}
     0.77 \pm 0.49_{f_i} \,; 
     &\quad E_\mathrm{cut} &= 20~\mathrm{MeV}, \\
     0.93 \pm 0.49_{f_i} \,; 
     &\quad E_\mathrm{cut} &= 50~\mathrm{MeV}, \\
     1.05 \pm 0.49_{f_i} \,; 
     &\quad E_\mathrm{cut} &= 100~\mathrm{MeV}, \\
     1.12 \pm 0.49_{f_i} \,; 
     &\quad E_\mathrm{cut} &= 150~\mathrm{MeV}. 
    \end{array}
   \right.
\end{align}

\begin{table}[t]
\centering
\renewcommand{\arraystretch}{1.15}
\setlength{\tabcolsep}{6pt}
\begin{tabular}{l|l|c}
\hline\rowcolor{\shadecolor{20}}
Parameter & Value & Reference \\
\hline
$G_F^{(\mu)}$ & $1.1663788\spac(6)\cdot 10^{-5}\,\mathrm{GeV}^{-2}$
 & \cite{ParticleDataGroup:2024cfk} \\
$\alpha(m_Z)^{-1}$ &$127.930\spac(8)$ & \cite{ParticleDataGroup:2024cfk} \\ 
$f_\pi$ & $130.2\spac(8)\,\mathrm{MeV}$ & \cite{FlavourLatticeAveragingGroupFLAG:2024oxs} \\
$f_B$ & $190.0\spac(1.3)\,\mathrm{MeV}$ & \cite{FlavourLatticeAveragingGroupFLAG:2024oxs} \\
$f_{B^\ast}/f_B$ & $0.951\spac(17)$ & \cite{Colquhoun:2015oha,Lubicz:2017asp} \\
$\abs{g_{BB^\ast\gamma}}$ &$1.45\spac(27)\,\mathrm{GeV^{-1}}$ & \cite{Pullin:2021ebn} \\
$\abs{g_{BB^\ast \pi}}$ & $0.56\spac(8)$ & \cite{Flynn:2015xna} \\
$\tau(B^-)$ & $1.637\spac(4)\,\mathrm{ps}$ & \cite{ParticleDataGroup:2024cfk} \\
\hline
\end{tabular}
\caption{\label{tab:inputs}
Numerical values for the input parameters entering our analysis. For the ratio $f_{B^\ast}/f_B$ we use the average of the available $N_f=2+1+1$ results from the HPQCD and ETMC collaborations. For a discussion on the input choices for $\abs{g_{BB^\ast\gamma}}$ and $\abs{g_{BB^\ast \pi}}$, see \cite{Cornella:2026lkp}.}
\end{table}

We now focus on the indirect contributions to the decay rate. For the channel involving the $B B^\ast\gamma$ coupling, we find 
\begin{align}
   \frac{\Gamma_{B^\ast \gamma}}{\Gamma_\mathrm{tree}}
   = \left\{ 
    \begin{array}{lll}
     (1.53\pm 0.57_{g_{BB^\ast\gamma}}\pm 0.05_{f_{B^\ast}/f_B})\cdot 10^{-7} \,; 
     &\quad E_\mathrm{cut} &= 20~\mathrm{MeV}, \\
     (3.17\pm 1.18_{g_{BB^\ast\gamma}}\pm 0.11_{f_{B^\ast}/f_B})\cdot 10^{-6} \,; 
     &\quad E_\mathrm{cut} &= 50~\mathrm{MeV}, \\
     (2.44\pm 0.91_{g_{BB^\ast\gamma}}\pm 0.09_{f_{B^\ast}/f_B})\cdot 10^{-5} \,; 
     &\quad E_\mathrm{cut} &= 100~\mathrm{MeV}, \\
     (7.28\pm 2.71_{g_{BB^\ast\gamma}}\pm 0.26_{f_{B^\ast}/f_B})\cdot 10^{-5} \,; 
     &\quad E_\mathrm{cut} &= 150~\mathrm{MeV}. 
    \end{array}
   \right.
\end{align}
The dominant uncertainty here comes from the coupling $g_{BB^\ast\gamma}$, but given the tiny overall size of the effect this has no impact on the final prediction. The contribution from the pion is only present for $E_\mathrm{cut}>m_{\pi^0}$. In the muon case it quickly overpowers the direct contribution once this threshold is crossed. For the tau mode, the overall impact is far more modest, as we demonstrate with the results
\begin{align}
   \frac{\Gamma_{B^\ast \pi}}{\Gamma_\mathrm{tree}}
   = \left\{ 
    \begin{array}{lll}
     (5.98\pm 1.71_{g_{BB^\ast\pi}}\pm 0.21_{f_{B^\ast}/f_B}\pm0.07_{f_\pi})\cdot 10^{-5} \,;  
     &\quad E_\mathrm{cut} &= 150~\mathrm{MeV}, \\
     (7.29\pm 2.08_{g_{BB^\ast\pi}}\pm 0.26_{f_{B^\ast}/f_B}\pm0.08_{f_\pi})\cdot 10^{-3} \,;
     &\quad E_\mathrm{cut} &= 250~\mathrm{MeV}, \\
     (2.87\pm 0.82_{g_{BB^\ast\pi}}\pm 0.10_{f_{B^\ast}/f_B}\pm0.04_{f_\pi})\cdot 10^{-2} \,; 
     &\quad E_\mathrm{cut} &= 350~\mathrm{MeV}. 
    \end{array}
   \right.
\end{align}
The rapid rise of this contribution is clearly visible, and a relevant contribution is obtained for values $E_{\rm cut}\gtrsim 250$\,MeV. Larger values of the veto energy would further enhance the pion mode, but would also extrapolate our prediction to a regime in which the EFT construction can no longer be considered reliable. Experimentally this contribution can likely be isolated by selecting events in which the invariant mass of photon pairs is close to the pion mass.

\begin{figure}
\centering
\includegraphics[scale=1.1]{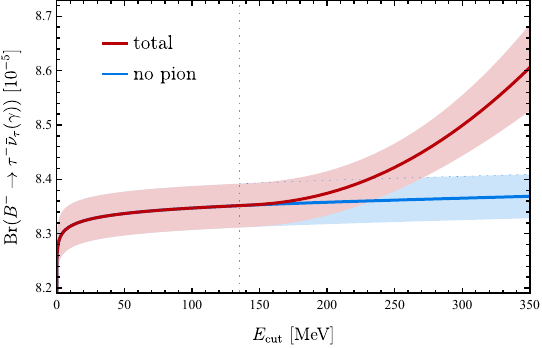}
\caption{$B^-\to\tau^-\spac\bar\nu_\tau(\gamma)$ branching ratio obtained with the central values for $|V_{ub}|$, $\tau(B^-)$ and $f_B$, both including (red band) and not including the pion contribution (blue band). The vertical dotted line shows the value of the pion mass, above which the latter contribution becomes relevant. The width of the bands indicates the theoretical uncertainties in the calculation of the structure-dependent QED corrections.}
\label{fig:rateplot}
\end{figure}

The above results display the size and uncertainties of the corrections normalized to the tree-level rate, the latter of which however contains additional uncertainties from $f_B$ and $|V_{ub}|$ that cancel out in the ratio we are taking. In practice, the uncertainty from $V_{ub}$ dominates the overall uncertainty by a wide margin. Therefore we show the value for the branching ratio for a general value $|V_{ub}|$, finding
\begin{align}\label{eq:BR}
\begin{aligned}
   \text{Br}\big(B^-\to\tau^-\spac\bar\nu_\tau(\gamma)\big) 
   &= \left[ \frac{\abs{V_{ub}}}{0.0036} \right]^2\! 
    \left[ \frac{\tau(B^-)}{1.637\,\text{ps}} \right]
    \left[ \frac{f_B}{190\,\text{MeV}} \right]^2 \cdot 10^{-5} \\
   &\quad\times \left\{
    \begin{array}{lll}
     8.32 \pm 0.04_{f_i} \,; &\quad E_\mathrm{cut} &= 20~\mathrm{MeV}, \\
     8.34 \pm 0.04_{f_i} \,; &\quad E_\mathrm{cut} &= 50~\mathrm{MeV}, \\
     8.35 \pm 0.04_{f_i} \,; &\quad E_\mathrm{cut} &= 150~\mathrm{MeV}, \\
     8.42 \pm 0.04_{f_i}\pm 0.02_{g_{BB^\ast\pi}} \,; &\quad E_\mathrm{cut} 
     &= 250~\mathrm{MeV}.
    \end{array}
   \right.
\end{aligned}
\end{align}
Here $f_B$ denotes the $B$-meson decay constant defined in pure QCD. Both the central values and the leading uncertainties stay almost constant over the shown range of $E_\mathrm{cut}$. Only for the largest value, the uncertainty from the $BB^\ast\pi$ coupling enters. We illustrate the behaviour of the branching ratio for a wide range of $E_\mathrm{cut}$ in Figure~\ref{fig:rateplot}. For reference, the blue line shows the central value when the pion contribution is neglected. 

\begin{figure}
\centering
\includegraphics[scale=1.1]{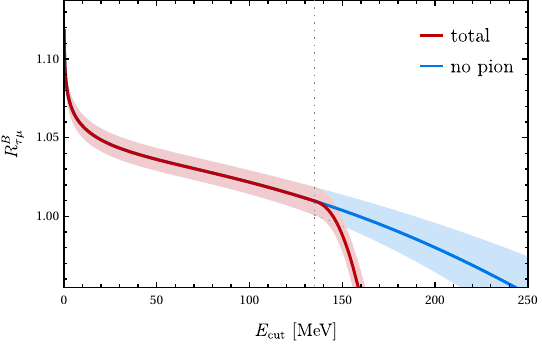}
\caption{Veto dependence of the LFU ratio $R_{\tau\mu}$ under the assumption that $E_\mathrm{cut}$ is chosen to be the same in the tau and muon channels. The red band shows the complete result, while the blue band assumes that the pion contribution has been removed. }
\label{fig:lfuplot}
\end{figure}

Finally, we compare the muon and tau channels directly by means of a lepton-flavor universality ratio, which we construct by dividing out the phase-space and chirality factors, i.e. 
\begin{align}\label{eq:R_LFU}
   R_{\tau\mu} 
   \equiv \frac{\Gamma\big(B^-\to\tau^-\spac\bar\nu_\tau(\gamma)\big)}%
               {\Gamma\big(B^-\to\mu^-\spac\bar\nu_\mu(\gamma)\big)}\,
    \frac{m_\mu^2}{m_\tau^2} \left( \frac{m_B^2-m_\mu^2}{m_B^2-m_\tau^2} \right)^2 .
\end{align}
In the absence of QED corrections, the quantity $R_{\tau\mu}$ defined this way is exactly equal to unity. While in principle there is no reason to choose the same value for the radiation veto in both channels, it is still interesting to study the dependence under the assumption of this choice. Combining the results from~\cite{Cornella:2026lkp} with those obtained here, we find
\begin{align}
   R_{\tau\mu} = \left\{ 
    \begin{array}{lll}
     1.049 \pm 0.007 \,; &\quad E_\mathrm{cut} &= 20~\mathrm{MeV}, \\
     1.036 \pm 0.007 \,; &\quad E_\mathrm{cut} &= 50~\mathrm{MeV}, \\
     1.022 \pm 0.008 \,; &\quad E_\mathrm{cut} &= 100~\mathrm{MeV}, \\
     1.004 \pm 0.010 \,; &\quad E_\mathrm{cut} &= 150~\mathrm{MeV}, \\
     0.950 \pm 0.024 \,; &\quad E_\mathrm{cut} &= 250~\mathrm{MeV}. 
    \end{array}
   \right.
\end{align}
In this result we have treated the pion contribution as a reducible background that can be experimentally identified and isolated, and hence it is not included in the calculation. As a consequence, the deviations from unity are modest across the whole range of veto energies. If, on the other hand, the pion background is included, the values for $E_\mathrm{cut}>m_\pi$ change to
\begin{align}
   R_{\tau\mu} = \left\{ 
    \begin{array}{lll}
     0.988 \pm 0.011 \,; &\quad E_\mathrm{cut} &= 150~\mathrm{MeV}, \\
     0.338 \pm 0.063 \,; &\quad E_\mathrm{cut} &= 250~\mathrm{MeV}, 
    \end{array}
   \right.
\end{align}
where the deviation from unity is driven mainly by the pion contribution to the muon channel. The uncertainties in the LFU ratio arise from non-universal QED corrections that do not cancel in the ratio, mostly given by the unknown hadronic parameters appearing in the predictions of the two decay rates. The complete energy dependence of $R_{\tau\mu}$ is shown in Figure~\ref{fig:lfuplot}. The ratio $R_{\tau\mu}$ was studied previously in~\cite{Dai:2021lei}, but the derivation in that study differs from ours in two ways. First, it treats both lepton flavors in the same EFT framework, more akin to our treatment of the muon channel. Second, the virtual corrections are computed assuming the meson to be a point-like heavy scalar, which ignores all structure-dependent corrections. As a result, the authors obtained slightly smaller values of $R_{\tau\mu}$ compared to ours, e.g.\ $R_{\tau\mu}=(1.031\pm 0.004)$ for $E_{\rm cut}=20$\,MeV.

\section{Conclusions}
\label{sec:conclusions}

We have presented a complete analysis of virtual and real QED corrections to the leptonic decay $B^-\to\tau^-\spac\bar\nu_\tau(\gamma)$, allowing for the emission of ultrasoft photons with energy less than a value $E_{\rm cut}$ in the $B$-meson rest frame. The leading logarithmic corrections have been resummed to all orders of perturbation theory. The calculation follows the effective field theory (EFT) strategy developed for the muon channel in~\cite{Cornella:2026lkp}, but the appropriate EFT below the $B$-meson mass is different in the present case. Since the tau lepton is not a highly boosted particle in the $B$-meson rest frame, ultrasoft-collinear modes are absent, and virtual electromagnetic corrections generate only local hadronic matrix elements, which we parameterize in terms of QED-induced $B^-\to\tau^-$ form factors.

The ``direct'' contributions to the decay are governed by the factorization formula \eqref{eq:totalQED}, which separates contributions from the hard scale $m_B$, the hadronic scale $\mu_0=1.5$\,GeV, and the ultrasoft scale set by the radiation veto $E_\mathrm{cut}$, and which includes the leading $1/m_\tau$-suppressed contributions. ``Indirect'' real-emission contributions arise at subleading order in EFT power counting and include $B^-\to B^{*-}\gamma$ and $B^-\to B^{*-}\pi^0$ transitions followed by the leptonic decay $B^{*-}\to\tau^-\spac\bar\nu_\tau$. In the second case the neutral pion decays into electromagnetic radiation with unity probability. Contrary to the muon case, these indirect contributions are found to be numerically very small and can be neglected as long as $E_{\rm cut}\ll\Lambda_{\rm QCD}$ -- a criterion required for the validity of the EFT construction. The suppression can be traced to the fact that the tau mass does not act as a chiral suppression, so that the compensation mechanism that enhances subleading effects in the muon channel is absent here.

We emphasize that the present analysis is not simply obtained by replacing the lepton mass in the muon-channel result. The two cases require different EFT constructions, leading to a structurally distinct factorization formula with different operator bases, anomalous dimensions, and relative importance of power-suppressed contributions. Overall, the QED corrections to the $B^-\to\tau^-\spac\bar\nu_\tau(\gamma)$ decay rate turn out to be small, at the level of a few percent, and dominated by the electroweak logarithms arising between $m_Z$ and $m_B$. The dependence on the radiation veto is mild across the range $E_\mathrm{cut}\in[20, 150]~\mathrm{MeV}$, reflecting the smallness of the velocity-dependent soft anomalous dimension $\gamma_\mathrm{soft}(v\cdot v_\tau)$ for the tau channel. Our final predictions for the corresponding branching ratio are given in \eqref{eq:BR} for some representative values of the veto energy, and shown in Figure~\ref{fig:rateplot} over a wider range of $E_{\rm cut}$ values. After factoring out $|V_{ub}|$ and the $B$-meson decay constant defined without QED effects, the QED-induced hadronic uncertainties entering through the form factors $f_i$ give rise to estimated theoretical uncertainties of 0.5\%, which could be reduced once these form factors are computed on the lattice. 

Our findings confirm that the $B^-\to\tau^-\spac\bar\nu_\tau(\gamma)$ decay rate is a theoretically clean observable, and that a precise extraction of $\abs{V_{ub}}$ from its measurement is feasible once the experimental precision improves. However, we emphasize that in order to use our clean theoretical predictions it will be necessary to lower the veto on real radiation to values $E_{\rm cut}\ll\Lambda_{\rm QCD}$. Finally, we have constructed the LFU ratio $R_{\tau\mu}$, normalized to remove the tree-level phase-space and helicity factors. In this quantity, the leading uncertainties arise from non-universal QED effects that do not cancel between the two channels. 

\subsection*{Acknowledgements}

This research has received funding from the Cluster of Excellence PRISMA${}^{++}$ (EXC 2118/2, Project ID 390831469) funded by the German Research Foundation (DFG) within the Germany Excellence Strategy, and from the European Research Council (ERC) under the European Union’s Horizon 2022 Research and Innovation Program (ERC Advanced Grant agreement No.~101097780, EFT4jets). Views and opinions expressed in this work are those of the authors only and do not necessarily reflect those of the European Union or the European Research Council Executive Agency. Neither the European Union nor the granting authority can be held responsible for them. 

\bibliographystyle{JHEP}
\bibliography{refs}

\end{document}